 \documentclass[preprint,prd,tightenlines, superscriptaddress]{revtex4-1}

\usepackage{graphicx} 
\usepackage{dcolumn}  
\usepackage{colordvi}
\usepackage{color}
\usepackage{epstopdf}
\usepackage{pstricks}
\usepackage{amssymb}
\usepackage{url}
\graphicspath{{ps}}
\usepackage{hyperref}
\usepackage{tabularx}
\usepackage{multirow}
\usepackage{units}
\usepackage{upgreek}
\usepackage{siunitx}
\usepackage{hyphenat}
\usepackage{subfigure}
\usepackage[italic]{hepnames}

\renewcommand{\PBzero}{\ensuremath{\HepParticle{\PB}{}{}^0}\xspace}
\renewcommand{\APBzero}{\ensuremath{\HepParticle{\APB}{}{}^0}\xspace}

\renewcommand{\APDzero}{\ensuremath{\HepParticle{\APD}{}{}^0}\xspace}
\renewcommand{\Pgpz}{\ensuremath{\HepParticle{\Pgp}{}{}^0}\xspace}
\renewcommand{\PDzero}{\ensuremath{\HepParticle{\PD}{}{}^0}\xspace}
\renewcommand{\PKzS}{\ensuremath{\HepParticle{\PK}{}{}^0_{\rm S}}\xspace}

\begin{document}

\def\belletwo {\it {Belle~II}}

\clubpenalty = 10000  
\widowpenalty = 10000 

\vspace*{-3\baselineskip}
\resizebox{!}{3cm}{\includegraphics{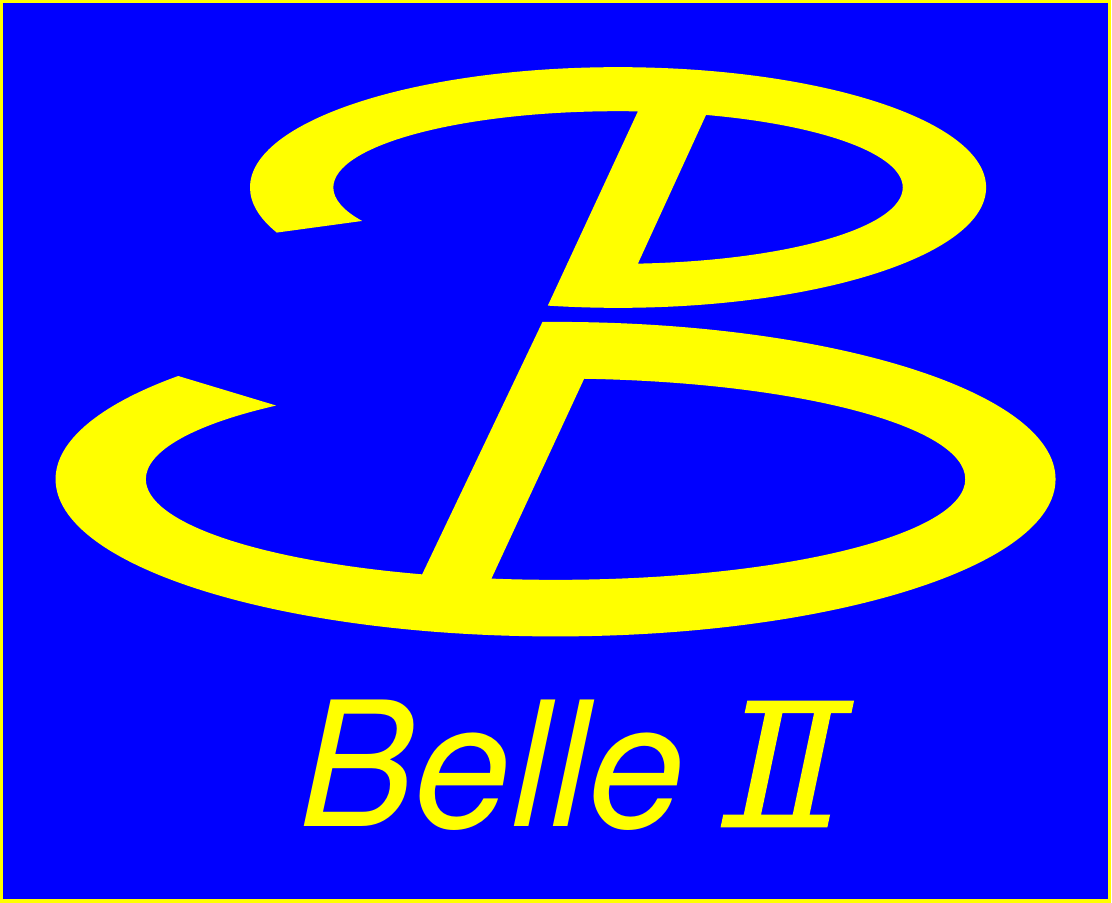}}

\vspace*{-5\baselineskip}
\begin{flushright}
BELLE2-CONF-PH-2021-002

\today
\end{flushright}

\quad\\[0.5cm]

\title {Measurements of branching fractions and CP-violating charge asymmetries in multibody charmless $B$ decays reconstructed in 2019--2020 Belle~II data}

\newcommand{\instCPPM}{Aix Marseille Universit\'{e}, CNRS/IN2P3, CPPM, 13288 Marseille, France}
\newcommand{\instBeihang}{Beihang University, Beijing 100191, China}
\newcommand{\instBNL}{Brookhaven National Laboratory, Upton, New York 11973, U.S.A.}
\newcommand{\instBINP}{Budker Institute of Nuclear Physics SB RAS, Novosibirsk 630090, Russian Federation}
\newcommand{\instCMU}{Carnegie Mellon University, Pittsburgh, Pennsylvania 15213, U.S.A.}
\newcommand{\instCinvestavIPN}{Centro de Investigacion y de Estudios Avanzados del Instituto Politecnico Nacional, Mexico City 07360, Mexico}
\newcommand{\instPrague}{Faculty of Mathematics and Physics, Charles University, 121 16 Prague, Czech Republic}
\newcommand{\instChiangMai}{Chiang Mai University, Chiang Mai 50202, Thailand}
\newcommand{\instChiba}{Chiba University, Chiba 263-8522, Japan}
\newcommand{\instChonnam}{Chonnam National University, Gwangju 61186, South Korea}
\newcommand{\instConacyt}{Consejo Nacional de Ciencia y Tecnolog\'{\i}a, Mexico City 03940, Mexico}
\newcommand{\instDESY}{Deutsches Elektronen--Synchrotron, 22607 Hamburg, Germany}
\newcommand{\instDuke}{Duke University, Durham, North Carolina 27708, U.S.A.}
\newcommand{\instITAR}{Institute of Theoretical and Applied Research (ITAR), Duy Tan University, Hanoi 100000, Vietnam}
\newcommand{\instRomaENEA}{ENEA Casaccia, I-00123 Roma, Italy}
\newcommand{\instEri}{Earthquake Research Institute, University of Tokyo, Tokyo 113-0032, Japan}
\newcommand{\instJuelich}{Forschungszentrum J\"{u}lich, 52425 J\"{u}lich, Germany}
\newcommand{\instFuJen}{Department of Physics, Fu Jen Catholic University, Taipei 24205, Taiwan}
\newcommand{\instFudan}{Key Laboratory of Nuclear Physics and Ion-beam Application (MOE) and Institute of Modern Physics, Fudan University, Shanghai 200443, China}
\newcommand{\instGoettingen}{II. Physikalisches Institut, Georg-August-Universit\"{a}t G\"{o}ttingen, 37073 G\"{o}ttingen, Germany}
\newcommand{\instGifu}{Gifu University, Gifu 501-1193, Japan}
\newcommand{\instSOKENDAI}{The Graduate University for Advanced Studies (SOKENDAI), Hayama 240-0193, Japan}
\newcommand{\instGyeongsang}{Gyeongsang National University, Jinju 52828, South Korea}
\newcommand{\instHanyang}{Department of Physics and Institute of Natural Sciences, Hanyang University, Seoul 04763, South Korea}
\newcommand{\instKEK}{High Energy Accelerator Research Organization (KEK), Tsukuba 305-0801, Japan}
\newcommand{\instJPARC}{J-PARC Branch, KEK Theory Center, High Energy Accelerator Research Organization (KEK), Tsukuba 305-0801, Japan}
\newcommand{\instHSE}{National Research University Higher School of Economics, Moscow 101000, Russian Federation}
\newcommand{\instIISER}{Indian Institute of Science Education and Research Mohali, SAS Nagar, 140306, India}
\newcommand{\instIITBhubaneswar}{Indian Institute of Technology Bhubaneswar, Satya Nagar 751007, India}
\newcommand{\instIITGuwahati}{Indian Institute of Technology Guwahati, Assam 781039, India}
\newcommand{\instIITHyderabad}{Indian Institute of Technology Hyderabad, Telangana 502285, India}
\newcommand{\instIITMadras}{Indian Institute of Technology Madras, Chennai 600036, India}
\newcommand{\instIndiana}{Indiana University, Bloomington, Indiana 47408, U.S.A.}
\newcommand{\instIHEPRussia}{Institute for High Energy Physics, Protvino 142281, Russian Federation}
\newcommand{\instHEPHYVienna}{Institute of High Energy Physics, Vienna 1050, Austria}
\newcommand{\instHiroshima}{Hiroshima University, Higashi-Hiroshima, Hiroshima 739-8530, Japan}
\newcommand{\instIHEPChina}{Institute of High Energy Physics, Chinese Academy of Sciences, Beijing 100049, China}
\newcommand{\instIPP}{Institute of Particle Physics (Canada), Victoria, British Columbia V8W 2Y2, Canada}
\newcommand{\instIOP}{Institute of Physics, Vietnam Academy of Science and Technology (VAST), Hanoi, Vietnam}
\newcommand{\instIFIC}{Instituto de Fisica Corpuscular, Paterna 46980, Spain}
\newcommand{\instFrascati}{INFN Laboratori Nazionali di Frascati, I-00044 Frascati, Italy}
\newcommand{\instNapoliINFN}{INFN Sezione di Napoli, I-80126 Napoli, Italy}
\newcommand{\instPadovaINFN}{INFN Sezione di Padova, I-35131 Padova, Italy}
\newcommand{\instPerugiaINFN}{INFN Sezione di Perugia, I-06123 Perugia, Italy}
\newcommand{\instPisaINFN}{INFN Sezione di Pisa, I-56127 Pisa, Italy}
\newcommand{\instRomaINFN}{INFN Sezione di Roma, I-00185 Roma, Italy}
\newcommand{\instRomaTreINFN}{INFN Sezione di Roma Tre, I-00146 Roma, Italy}
\newcommand{\instTorinoINFN}{INFN Sezione di Torino, I-10125 Torino, Italy}
\newcommand{\instTriesteINFN}{INFN Sezione di Trieste, I-34127 Trieste, Italy}
\newcommand{\instJAEA}{Advanced Science Research Center, Japan Atomic Energy Agency, Naka 319-1195, Japan}
\newcommand{\instMainz}{Johannes Gutenberg-Universit\"{a}t Mainz, Institut f\"{u}r Kernphysik, D-55099 Mainz, Germany}
\newcommand{\instGiessen}{Justus-Liebig-Universit\"{a}t Gie\ss{}en, 35392 Gie\ss{}en, Germany}
\newcommand{\instKarlsruhe}{Institut f\"{u}r Experimentelle Teilchenphysik, Karlsruher Institut f\"{u}r Technologie, 76131 Karlsruhe, Germany}
\newcommand{\instISU}{Iowa State University, Ames, Iowa 50011, U.S.A.}
\newcommand{\instKitasato}{Kitasato University, Sagamihara 252-0373, Japan}
\newcommand{\instKISTI}{Korea Institute of Science and Technology Information, Daejeon 34141, South Korea}
\newcommand{\instKoreaUnivKU}{Korea University, Seoul 02841, South Korea}
\newcommand{\instKSU}{Kyoto Sangyo University, Kyoto 603-8555, Japan}
\newcommand{\instKyungpook}{Kyungpook National University, Daegu 41566, South Korea}
\newcommand{\instLPI}{P.N. Lebedev Physical Institute of the Russian Academy of Sciences, Moscow 119991, Russian Federation}
\newcommand{\instLNNU}{Liaoning Normal University, Dalian 116029, China}
\newcommand{\instLMU}{Ludwig Maximilians University, 80539 Munich, Germany}
\newcommand{\instLuther}{Luther College, Decorah, Iowa 52101, U.S.A.}
\newcommand{\instMNITJaipur}{Malaviya National Institute of Technology Jaipur, Jaipur 302017, India}
\newcommand{\instMPP}{Max-Planck-Institut f\"{u}r Physik, 80805 M\"{u}nchen, Germany}
\newcommand{\instMPGHLL}{Semiconductor Laboratory of the Max Planck Society, 81739 M\"{u}nchen, Germany}
\newcommand{\instMcGill}{McGill University, Montr\'{e}al, Qu\'{e}bec, H3A 2T8, Canada}
\newcommand{\instMEPhI}{Moscow Physical Engineering Institute, Moscow 115409, Russian Federation}
\newcommand{\instNagoya}{Graduate School of Science, Nagoya University, Nagoya 464-8602, Japan}
\newcommand{\instNagoyaIAR}{Institute for Advanced Research, Nagoya University, Nagoya 464-8602, Japan}
\newcommand{\instNagoyaKMI}{Kobayashi-Maskawa Institute, Nagoya University, Nagoya 464-8602, Japan}
\newcommand{\instNaraWu}{Nara Women's University, Nara 630-8506, Japan}
\newcommand{\instNTUTaiwan}{Department of Physics, National Taiwan University, Taipei 10617, Taiwan}
\newcommand{\instNUUTaiwan}{National United University, Miao Li 36003, Taiwan}
\newcommand{\instKrakow}{H. Niewodniczanski Institute of Nuclear Physics, Krakow 31-342, Poland}
\newcommand{\instNiigata}{Niigata University, Niigata 950-2181, Japan}
\newcommand{\instNSU}{Novosibirsk State University, Novosibirsk 630090, Russian Federation}
\newcommand{\instOkinawa}{Okinawa Institute of Science and Technology, Okinawa 904-0495, Japan}
\newcommand{\instOsakaCity}{Osaka City University, Osaka 558-8585, Japan}
\newcommand{\instRCNP}{Research Center for Nuclear Physics, Osaka University, Osaka 567-0047, Japan}
\newcommand{\instPNNL}{Pacific Northwest National Laboratory, Richland, Washington 99352, U.S.A.}
\newcommand{\instPanjab}{Panjab University, Chandigarh 160014, India}
\newcommand{\instPanjabPAU}{Punjab Agricultural University, Ludhiana 141004, India}
\newcommand{\instRIKENMSL}{Meson Science Laboratory, Cluster for Pioneering Research, RIKEN, Saitama 351-0198, Japan}
\newcommand{\instSeoul}{Seoul National University, Seoul 08826, South Korea}
\newcommand{\instSPU}{Showa Pharmaceutical University, Tokyo 194-8543, Japan}
\newcommand{\instSoochow}{Soochow University, Suzhou 215006, China}
\newcommand{\instSoongsil}{Soongsil University, Seoul 06978, South Korea}
\newcommand{\instLjubljanaJSI}{J. Stefan Institute, 1000 Ljubljana, Slovenia}
\newcommand{\instKyiv}{Taras Shevchenko National Univ. of Kiev, Kiev, Ukraine}
\newcommand{\instTata}{Tata Institute of Fundamental Research, Mumbai 400005, India}
\newcommand{\instTUM}{Department of Physics, Technische Universit\"{a}t M\"{u}nchen, 85748 Garching, Germany}
\newcommand{\instTelAviv}{Tel Aviv University, School of Physics and Astronomy, Tel Aviv, 69978, Israel}
\newcommand{\instToho}{Toho University, Funabashi 274-8510, Japan}
\newcommand{\instTohoku}{Department of Physics, Tohoku University, Sendai 980-8578, Japan}
\newcommand{\instTitech}{Tokyo Institute of Technology, Tokyo 152-8550, Japan}
\newcommand{\instTokyoMetropolitan}{Tokyo Metropolitan University, Tokyo 192-0397, Japan}
\newcommand{\instUAS}{Universidad Autonoma de Sinaloa, Sinaloa 80000, Mexico}
\newcommand{\instNapoliUNIV}{Dipartimento di Scienze Fisiche, Universit\`{a} di Napoli Federico II, I-80126 Napoli, Italy}
\newcommand{\instPadovaUNIV}{Dipartimento di Fisica e Astronomia, Universit\`{a} di Padova, I-35131 Padova, Italy}
\newcommand{\instPerugiaUNIV}{Dipartimento di Fisica, Universit\`{a} di Perugia, I-06123 Perugia, Italy}
\newcommand{\instPisaUNIV}{Dipartimento di Fisica, Universit\`{a} di Pisa, I-56127 Pisa, Italy}
\newcommand{\instRomaTreUNIV}{Dipartimento di Matematica e Fisica, Universit\`{a} di Roma Tre, I-00146 Roma, Italy}
\newcommand{\instTorinoUNIV}{Dipartimento di Fisica, Universit\`{a} di Torino, I-10125 Torino, Italy}
\newcommand{\instTriesteUNIV}{Dipartimento di Fisica, Universit\`{a} di Trieste, I-34127 Trieste, Italy}
\newcommand{\instMontreal}{Universit\'{e} de Montr\'{e}al, Physique des Particules, Montr\'{e}al, Qu\'{e}bec, H3C 3J7, Canada}
\newcommand{\instIJCLab}{Universit\'{e} Paris-Saclay, CNRS/IN2P3, IJCLab, 91405 Orsay, France}
\newcommand{\instIPHC}{Universit\'{e} de Strasbourg, CNRS, IPHC, UMR 7178, 67037 Strasbourg, France}
\newcommand{\instAdelaide}{Department of Physics, University of Adelaide, Adelaide, South Australia 5005, Australia}
\newcommand{\instBonn}{University of Bonn, 53115 Bonn, Germany}
\newcommand{\instUBC}{University of British Columbia, Vancouver, British Columbia, V6T 1Z1, Canada}
\newcommand{\instCincinnati}{University of Cincinnati, Cincinnati, Ohio 45221, U.S.A.}
\newcommand{\instFlorida}{University of Florida, Gainesville, Florida 32611, U.S.A.}
\newcommand{\instHawaii}{University of Hawaii, Honolulu, Hawaii 96822, U.S.A.}
\newcommand{\instHeidelberg}{University of Heidelberg, 68131 Mannheim, Germany}
\newcommand{\instLjubljanaUniLJ}{Faculty of Mathematics and Physics, University of Ljubljana, 1000 Ljubljana, Slovenia}
\newcommand{\instLouisville}{University of Louisville, Louisville, Kentucky 40292, U.S.A.}
\newcommand{\instMalaya}{National Centre for Particle Physics, University Malaya, 50603 Kuala Lumpur, Malaysia}
\newcommand{\instLjubljanaUM}{Faculty of Chemistry and Chemical Engineering, University of Maribor, 2000 Maribor, Slovenia}
\newcommand{\instMelbourne}{School of Physics, University of Melbourne, Victoria 3010, Australia}
\newcommand{\instMississippi}{University of Mississippi, University, Mississippi 38677, U.S.A.}
\newcommand{\instUOM}{University of Miyazaki, Miyazaki 889-2192, Japan}
\newcommand{\instPittsburgh}{University of Pittsburgh, Pittsburgh, Pennsylvania 15260, U.S.A.}
\newcommand{\instUSTC}{University of Science and Technology of China, Hefei 230026, China}
\newcommand{\instSAlabama}{University of South Alabama, Mobile, Alabama 36688, U.S.A.}
\newcommand{\instSCarolina}{University of South Carolina, Columbia, South Carolina 29208, U.S.A.}
\newcommand{\instSydney}{School of Physics, University of Sydney, New South Wales 2006, Australia}
\newcommand{\instUTokyo}{Department of Physics, University of Tokyo, Tokyo 113-0033, Japan}
\newcommand{\instIPMU}{Kavli Institute for the Physics and Mathematics of the Universe (WPI), University of Tokyo, Kashiwa 277-8583, Japan}
\newcommand{\instVictoria}{University of Victoria, Victoria, British Columbia, V8W 3P6, Canada}
\newcommand{\instVPI}{Virginia Polytechnic Institute and State University, Blacksburg, Virginia 24061, U.S.A.}
\newcommand{\instWayneState}{Wayne State University, Detroit, Michigan 48202, U.S.A.}
\newcommand{\instYamagata}{Yamagata University, Yamagata 990-8560, Japan}
\newcommand{\instYerevan}{Alikhanyan National Science Laboratory, Yerevan 0036, Armenia}
\newcommand{\instYonsei}{Yonsei University, Seoul 03722, South Korea}
\newcommand{\instZZU}{Zhengzhou University, Zhengzhou 450001, China}
\affiliation{\instCPPM}
\affiliation{\instBeihang}
\affiliation{\instBNL}
\affiliation{\instBINP}
\affiliation{\instCMU}
\affiliation{\instCinvestavIPN}
\affiliation{\instPrague}
\affiliation{\instChiangMai}
\affiliation{\instChiba}
\affiliation{\instChonnam}
\affiliation{\instConacyt}
\affiliation{\instDESY}
\affiliation{\instDuke}
\affiliation{\instITAR}
\affiliation{\instRomaENEA}
\affiliation{\instEri}
\affiliation{\instJuelich}
\affiliation{\instFuJen}
\affiliation{\instFudan}
\affiliation{\instGoettingen}
\affiliation{\instGifu}
\affiliation{\instSOKENDAI}
\affiliation{\instGyeongsang}
\affiliation{\instHanyang}
\affiliation{\instKEK}
\affiliation{\instJPARC}
\affiliation{\instHSE}
\affiliation{\instIISER}
\affiliation{\instIITBhubaneswar}
\affiliation{\instIITGuwahati}
\affiliation{\instIITHyderabad}
\affiliation{\instIITMadras}
\affiliation{\instIndiana}
\affiliation{\instIHEPRussia}
\affiliation{\instHEPHYVienna}
\affiliation{\instHiroshima}
\affiliation{\instIHEPChina}
\affiliation{\instIPP}
\affiliation{\instIOP}
\affiliation{\instIFIC}
\affiliation{\instFrascati}
\affiliation{\instNapoliINFN}
\affiliation{\instPadovaINFN}
\affiliation{\instPerugiaINFN}
\affiliation{\instPisaINFN}
\affiliation{\instRomaINFN}
\affiliation{\instRomaTreINFN}
\affiliation{\instTorinoINFN}
\affiliation{\instTriesteINFN}
\affiliation{\instJAEA}
\affiliation{\instMainz}
\affiliation{\instGiessen}
\affiliation{\instKarlsruhe}
\affiliation{\instISU}
\affiliation{\instKitasato}
\affiliation{\instKISTI}
\affiliation{\instKoreaUnivKU}
\affiliation{\instKSU}
\affiliation{\instKyungpook}
\affiliation{\instLPI}
\affiliation{\instLNNU}
\affiliation{\instLMU}
\affiliation{\instLuther}
\affiliation{\instMNITJaipur}
\affiliation{\instMPP}
\affiliation{\instMPGHLL}
\affiliation{\instMcGill}
\affiliation{\instMEPhI}
\affiliation{\instNagoya}
\affiliation{\instNagoyaIAR}
\affiliation{\instNagoyaKMI}
\affiliation{\instNaraWu}
\affiliation{\instNTUTaiwan}
\affiliation{\instNUUTaiwan}
\affiliation{\instKrakow}
\affiliation{\instNiigata}
\affiliation{\instNSU}
\affiliation{\instOkinawa}
\affiliation{\instOsakaCity}
\affiliation{\instRCNP}
\affiliation{\instPNNL}
\affiliation{\instPanjab}
\affiliation{\instPanjabPAU}
\affiliation{\instRIKENMSL}
\affiliation{\instSeoul}
\affiliation{\instSPU}
\affiliation{\instSoochow}
\affiliation{\instSoongsil}
\affiliation{\instLjubljanaJSI}
\affiliation{\instKyiv}
\affiliation{\instTata}
\affiliation{\instTUM}
\affiliation{\instTelAviv}
\affiliation{\instToho}
\affiliation{\instTohoku}
\affiliation{\instTitech}
\affiliation{\instTokyoMetropolitan}
\affiliation{\instUAS}
\affiliation{\instNapoliUNIV}
\affiliation{\instPadovaUNIV}
\affiliation{\instPerugiaUNIV}
\affiliation{\instPisaUNIV}
\affiliation{\instRomaTreUNIV}
\affiliation{\instTorinoUNIV}
\affiliation{\instTriesteUNIV}
\affiliation{\instMontreal}
\affiliation{\instIJCLab}
\affiliation{\instIPHC}
\affiliation{\instAdelaide}
\affiliation{\instBonn}
\affiliation{\instUBC}
\affiliation{\instCincinnati}
\affiliation{\instFlorida}
\affiliation{\instHawaii}
\affiliation{\instHeidelberg}
\affiliation{\instLjubljanaUniLJ}
\affiliation{\instLouisville}
\affiliation{\instMalaya}
\affiliation{\instLjubljanaUM}
\affiliation{\instMelbourne}
\affiliation{\instMississippi}
\affiliation{\instUOM}
\affiliation{\instPittsburgh}
\affiliation{\instUSTC}
\affiliation{\instSAlabama}
\affiliation{\instSCarolina}
\affiliation{\instSydney}
\affiliation{\instUTokyo}
\affiliation{\instIPMU}
\affiliation{\instVictoria}
\affiliation{\instVPI}
\affiliation{\instWayneState}
\affiliation{\instYamagata}
\affiliation{\instYerevan}
\affiliation{\instYonsei}
\affiliation{\instZZU}
  \author{F.~Abudin{\'e}n}\affiliation{\instTriesteINFN} 
  \author{I.~Adachi}\affiliation{\instKEK}\affiliation{\instSOKENDAI} 
  \author{R.~Adak}\affiliation{\instFudan} 
  \author{K.~Adamczyk}\affiliation{\instKrakow} 
  \author{P.~Ahlburg}\affiliation{\instBonn} 
  \author{J.~K.~Ahn}\affiliation{\instKoreaUnivKU} 
  \author{H.~Aihara}\affiliation{\instUTokyo} 
  \author{N.~Akopov}\affiliation{\instYerevan} 
  \author{A.~Aloisio}\affiliation{\instNapoliUNIV}\affiliation{\instNapoliINFN} 
  \author{F.~Ameli}\affiliation{\instRomaINFN} 
% ...
\collaboration{Belle II Collaboration}


\begin{abstract}
We report on measurements of branching fractions~($\mathcal{B}$) and CP-violating charge asymmetries~($\mathcal{A}_{\rm CP}$) of multibody charmless $B$ decays reconstructed by the Belle~II experiment at the SuperKEKB electron-positron collider.  We use a sample of collisions collected in 2019 and 2020 at the $\Upsilon(4S)$ resonance and  corresponding to $62.8$\,fb$^{-1}$ of integrated luminosity. We use simulation to determine optimized event selections. The $\Delta E$ and $M_{\rm bc}$ distributions of the resulting samples are fit to determine signal yields of approximately 690, 840, and 380 decays for the channels 
\mbox{$B^+ \to K^+K^-K^+$}, \mbox{$B^+ \to K^+\pi^-\pi^+$}, and  \mbox{$B^0 \to K^+\pi^-\pi^0$}, respectively. These yields are corrected for efficiencies determined from simulation and control data samples to obtain
\begin{center}
$\mathcal{B}(B^+ \to K^+K^-K^+) = [35.8 \pm 1.6(\rm stat) \pm 1.4 (\rm syst)]\times 10^{-6}$, 
\end{center}
\begin{center}
$\mathcal{B}(B^+ \to K^+\pi^-\pi^+) = [67.0 \pm 3.3 (\rm stat)\pm 2.3 (\rm syst)]\times 10^{-6}$,
\end{center}
\begin{center}
$\mathcal{B}(B^0 \to K^+\pi^-\pi^0) = [38.1 \pm 3.5 (\rm stat)\pm 3.9 (\rm syst)]\times 10^{-6}$, 
\end{center}
\begin{center}
$\mathcal{A}_{\rm CP}(B^+ \to K^+K^-K^+) = -0.103 \pm 0.042(\rm stat) \pm 0.020 (\rm syst)$,
\end{center}
\begin{center}
$\mathcal{A}_{\rm CP}(B^+ \to K^+\pi^-\pi^+) = -0.010 \pm 0.050 (\rm stat)\pm 0.021(\rm syst)$, \mbox{and}
\end{center}
\begin{center}
$\mathcal{A}_{\rm CP}(B^0 \to K^+\pi^-\pi^0) = 0.207 \pm 0.088 (\rm stat)\pm 0.011(\rm syst)$.
\end{center}
Results are consistent with previous measurements and demonstrate detector performance comparable with the best Belle results.
 
\end{abstract}


\maketitle

{\renewcommand{\thefootnote}{\fnsymbol{footnote}}}
\setcounter{footnote}{0}



\section{Introduction and motivation}

The study of multibody charmless $B$ decays has recently attracted significant attention in the worldwide flavor program.
The phenomenology of the interplay between weak- and strong-interaction dynamics in these decays is enriched by the amplitude structure accessible through the Dalitz plot. Previous measurements exposed large local charge-parity~(CP) violating asymmetries~\cite{Lees:2012kxa, Aubert:2008bj} whose interpretation prompted significant activity~\cite{PhysRevLett.112.011801, JHEP10(2017)117, PhysRevD.94.094015}.
The Belle  II  physics  program,  featuring the  {\it unique}  capability  of  studying  jointly,  and within a consistent experimental environment, all relevant final states is particularly promising to achieve a consistent global picture.

The Belle~II detector, complete with its silicon tracker, started its physics operations at the SuperKEKB asymmetric-energy collider on March 11, 2019.  The sample of electron-positron collisions used in this work corresponds to an integrated luminosity of $62.8\,\si{fb^{-1}}$~\cite{Abudinen:2019osb} and was collected at the $\Upsilon(4{\rm S})$~resonance as of July 1, 2020.
We report on measurements of branching fractions and CP-violating charge asymmetries in  multibody charmless decays at Belle~II updated with more data, additional channels, and more refined analyses compared with previous results ~\cite{Benedikt:2019,CharmlessMoriond:2020}. 
 The target decay modes are \mbox{$B^+ \to K^+K^-K^+$}, \mbox{$B^+ \to K^+\pi^-\pi^+$}, and \mbox{$B^0 \to K^+\pi^-\pi^0$}.
Charge-conjugate processes are implied in what follows unless otherwise stated. Analysis improvements over our previous results are the inclusion of the $B^0 \to K^+\pi^-\pi^0$ channel; improved sample-composition determinations, which are now based on simultaneous fits to the energy-difference and beam-constrained-mass distributions; and a refined treatment of peaking-background contributions, signal efficiencies, and systematic uncertainties. \par
All analysis procedures are developed and finalized in simulated data prior to be applied to the experimental signal sample. The $B^+ \to K^+K^-K^+$ and $B^+ \to K^+\pi^-\pi^+$ decays are subjected to an additional validation on half of the experimental data sample. Optimized event selections are determined using simulated and control sample data. The composition of resulting samples is then determined using fits to the following observables:
\begin{itemize}
 \item the beam-energy-constrained mass $M_{\rm bc} \equiv \sqrt{s/(4c^4) - (p^{*}_B/c)^2}$, which is the $B$ candidate mass with $B$ energy replaced by the (more precisely known) half of the center-of-mass energy, and discriminates fully reconstructed $B$ decays from $e^+e^- \to q\bar{q}$ background events, where $q$ is any quark lighter than the $b$ quark.
    \item the energy difference $\Delta E \equiv E^{*}_{B} - \sqrt{s}/2$ between the total energy of the reconstructed $B$ candidate and half of the center-of-mass energy, both in the $\Upsilon(4S)$ frame, which provides additional discrimination between correctly and incorrectly reconstructed $B$ decays.
   
\end{itemize}

\section{The Belle~II detector}
Belle~II is a nearly $4\pi$ particle-physics spectrometer~\cite{Kou:2018nap, Abe:2010sj}, designed to reconstruct the products of electron-positron collisions produced by the SuperKEKB asymmetric-energy collider~\cite{Akai:2018mbz}, located at the KEK laboratory in Tsukuba, Japan. Belle~II comprises several subdetectors arranged around the interaction space-point in a cylindrical geometry. The innermost subdetector is the vertex detector, which uses position-sensitive silicon layers to sample the trajectories of charged particles (tracks) in the vicinity of the interaction region to extrapolate the decay positions of their long-lived parent particles. The vertex detector includes two inner layers of pixel sensors and four outer layers of double-sided microstrip sensors. The second pixel layer is currently incomplete and covers only one sixth of azimuthal angle. Charged-particle momenta and charges are measured by a large-radius, helium-ethane, small-cell central drift chamber, which also offers charged-particle-identification information through a measurement of particles' energy-loss by specific ionization. A Cherenkov-light angle and time-of-propagation detector surrounding the chamber provides charged-particle identification in the central detector volume, supplemented by proximity-focusing, aerogel, ring-imaging Cherenkov detectors in the forward region. A CsI(Tl)-crystal electromagnetic calorimeter allows for energy measurements of electrons and photons.  A solenoid surrounding the calorimeter generates a uniform axial 1.5\,T magnetic field filling its inner volume. Layers of plastic scintillator and resistive-plate chambers, interspersed between the
magnetic flux-return iron plates, allow for identification of $K^0_{\rm L}$ and muons.
The subdetectors most relevant for this work are the silicon vertex detector, the drift chamber, the particle-identification detectors, and the electromagnetic calorimeter.

\section{Selection and reconstruction}
\label{sec:selection}

We reconstruct the three-body decays
\begin{itemize}
    \item $B^+ \to K^+\, K^-\, K^+$,
    \item $B^+ \to K^+\, \pi^+\, \pi^-$.
    \item $B^0 \to K^+ \pi^- \pi^0(\to \gamma\gamma)$.
\end{itemize}
In addition, we use the control channels
\begin{itemize}
\item \mbox{$B^+ \to \overline{D}^0 (\to K^+ \pi^- \pi^0(\to \gamma\gamma))\, \pi^+$}, 
\item \mbox{$B^+ \to \overline{D}^0 (\to K^+ \pi^-)\, \pi^+$}, 
\item \mbox{$B^0 \to D^{*-}(\to \overline{D}^0 (\to K^+ \pi^- \pi^0(\to \gamma\gamma))\, \pi^-)\, \pi^+$}, 
\item \mbox{$B^0 \to D^{*-}(\to \overline{D}^0 (\to K^+ \pi^-)\, \pi^-)\, \pi^+$}, 
\item \mbox{$D^+ \to \PKzS(\to \pi^+\pi^-)\pi^+$},
\item \mbox{$\PDzero \to K^-\pi^+$}, 
\end{itemize}
for validation of continuum-suppression discriminating variables; optimization of the $\pi^0$ selection and determination of its efficiency; assessment of data-simulation discrepancies in the particle-identification quantities; and determination of instrumental asymmetries. 


\subsection{Simulated and experimental data}
We use simulated generic $e^+e^-$-collision data to optimize the event selection and determine composition-fit models for nonsignal components.  We use signal-only simulated data to model relevant signal features for fits and determine selection efficiencies.
Simplified simulated experiments obtained by randomly sampling the likelihood used for the sample-composition fit are used to assess systematic uncertainties on modeling. Generic simulation consists of Monte Carlo samples that include $e^+e^-\to B^0\overline{B}^0$, $B^+B^-$, $u\bar{u}$, $d\bar{d}$, $c\bar{c}$, and $s\bar{s}$ processes in realistic proportions and corresponding in size to 2--10  times the $\Upsilon$(4S) data. In addition, $6\times 10^5$\mbox{$B^+ \to K^+ K^- K^+$} and \mbox{$B^+ \to K^+ \pi^- \pi^+$} decays are generated along with $2\times 10^6$ \mbox{$B^0 \to K^+ \pi^- \pi^0$} decays, assuming a simplified Dalitz structure win which major resonances are present but are not interfering~\cite{Ryd:2005zz}. \par 
We use all 2019--2020  $\Upsilon$(4S) good-quality experimental data collected up to July~1, 2020, which correspond to an integrated luminosity of $62.8\,\si{fb^{-1}}$. All events are required to satisfy loose data-skim selection criteria, based on total energy and charged-particle multiplicity in the event, targeted at reducing sample sizes to a manageable level with negligible impact on signal efficiency. All data are processed using the Belle~II analysis software~\cite{Kuhr:2018lps}.

\subsection{Reconstruction and baseline selection}
We form final-state particle candidates by applying loose initial selection criteria and then combining candidates in kinematic fits consistent with the topologies of the desired decays to reconstruct intermediate states and $B$ candidates. \par We reconstruct charged-pion and -kaon candidates using inclusive charged-particle selections restricted to the full polar-angle acceptance in the central drift chamber ($\SI{17}{\degree}<\theta<\SI{150}{\degree}$) and to loose ranges of displacement from the nominal interaction space-point ($|dr|<\SI{0.5}{cm}$ radial with respect to the beam axis and $|dz|<\SI{3}{cm}$ longitudinal) to reduce beam-background-induced tracks, which do not originate from the interaction region.
We reconstruct neutral-pion candidates by combining pairs of photons with energies greater than about $20$\,MeV restricted in diphoton mass and excluding extreme helicity-angle values  to suppress combinatorial background from soft photons. In addition, a binary boosted decision-tree classifier is trained on calorimeter variables to distinguish photons coming from $B$ decays from those associated with Bhabha processes. The mass of each $\pi^0$ candidate is constrained to its known value~\cite{Zyla:2020zbs} in subsequent kinematic fits. The momentum is required to exceed $0.5$ GeV/$c$ to maximize ${\rm S}/\sqrt{{\rm S}+{\rm B}}$, where ${\rm S}$ and ${\rm B}$ are simulated signal and background yields in the signal region. The resulting $K^+$, $\pi^+$, and $\pi^0$ candidates are combined through simultaneous kinematic fits of the entire decay chain into each of our target signal channels, consistent with the desired topology. In addition, we reconstruct the vertex of the accompanying tag-side $B$~mesons using  all tracks in the tag-side and identify the flavor, which is used as  input to the continuum-background discriminator, using a category-based flavor tagger~\cite{Abudinen:2018}.
The reconstruction of the control channels is conceptually similar.
\par
The resulting samples include contributions from signal events, self-cross-feed (i.e., incorrectly reconstructed candidates in signal events), continuum background, and peaking backgrounds, that is, misreconstructed events clustering in the signal region.
We use simulation to identify and suppress contamination from peaking backgrounds.
\subsection{Charmed-background vetoes}
Dominant \mbox{$B^0 \to \overline{D}^0(\to K^+K^-)K^+$}, \mbox{$B^0 \to \eta_c(\to K^+K^-)K^+$}, and \mbox{$B^0 \to \chi_{c1}(\to K^+K^-)K^+$} contributions to the  
$B^0 \to K^+K^-K^+$ sample are suppressed by excluding the two-body mass ranges ranges \mbox{$1.84 < m(K^+K^-) < 1.88$\,GeV/$c^2$}, \mbox{$2.94 < m(K^+K^-) < 3.05$\,GeV/$c^2$}, and \mbox{$3.50 < m(K^+K^-) < 3.54$ GeV/$c^2$}, respectively.  
The $B^+ \to K^+\pi^-\pi^+$ channel is contaminated by the charmed intermediate states $B^+ \to \overline{D}^0(\to h^+h^{(')-})\pi^+$ (where $h$ and $h^{'}$ are either kaons or pions),  $B^+ \to \eta_c(\to \pi^+\pi^-)K^+$, \mbox{$B^+ \to \chi_{c1}(\to \pi^+\pi^-)K^+$}, and $B^+ \to \eta_c(2S)(\to \pi^+\pi^-)K^+$, and intermediate resonances decaying to muons misidentified as pions $B^+ \to J/\psi(\to \mu^+\mu^-)K^+$ and $B^+ \to \psi(2S)(\to \mu^+\mu^-)K^+$.
These are suppressed by excluding the two-body mass ranges $1.8 < m(h^+h^{(')-}) < 1.92$ GeV/$c^2$, $2.93 < m(\pi\pi) < 3.15$ GeV/$c^2$, $3.45 < m(\pi\pi) < 3.525$ GeV/$c^2$, $3.62 < m(\pi\pi) < 3.665$ GeV/$c^2$, $3.67 < m(\pi\pi) < 3.72$ GeV/$c^2$. The $B^0 \to K^+\pi^-\pi^0$ candidates are contaminated by $B$ decays proceeding through intermediate $D$ meson decays, including  $B^{+}\rightarrow \overline{D}^{0}(\to h^{+}h^{(')-})\rho^{+}(\pi^{0}...)$, $B^{+}\rightarrow \overline{D}^{0}(\to h^{+}h^{(')-})\pi^{0}$, and  $B^{0}\rightarrow \overline{D}^{*}(2007)^{0}(\to \overline{D}^{0}(h^{+}h^{(')-})...)\pi^{0}$ decays, where $h$ and $h^{'}$ are kaons or pions that could be either properly identified or misidentified. We veto candidates with kaon-pion mass between $1.8$ and $1.92$\,GeV/$c^{2}$.


\subsection{Continuum suppression}
The main challenge in observing significant charmless signals is the large contamination from continuum background. We use a binary boosted decision-tree classifier that nonlinearly combines 39 variables known to provide statistical discrimination between $B$-meson signals and continuum and to be loosely correlated, or uncorrelated,  with $\Delta E$ and $M_{\rm bc}$. The variables include quantities associated to event topology (global and signal-only angular configurations), flavor-tagger information, vertex separation and uncertainty information, and kinematic-fit quality information. We train the classifier to identify statistically significant signal and background features using simulated samples. 

\begin{figure}[h!]
 \centering
    \centering
    \subfigure{\includegraphics[width=0.49\textwidth]{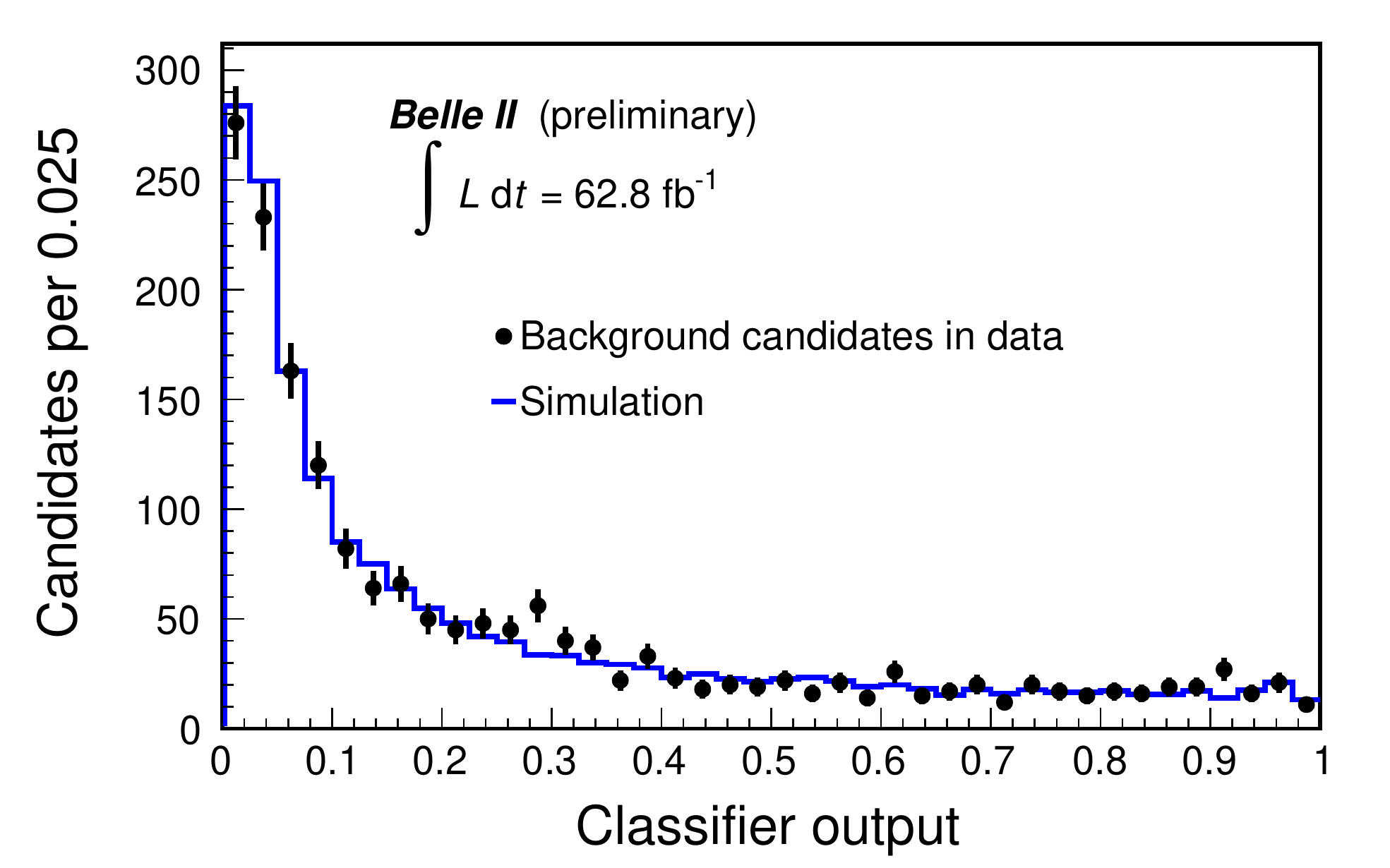}} \hfill
    \subfigure{\includegraphics[width=0.49\textwidth]{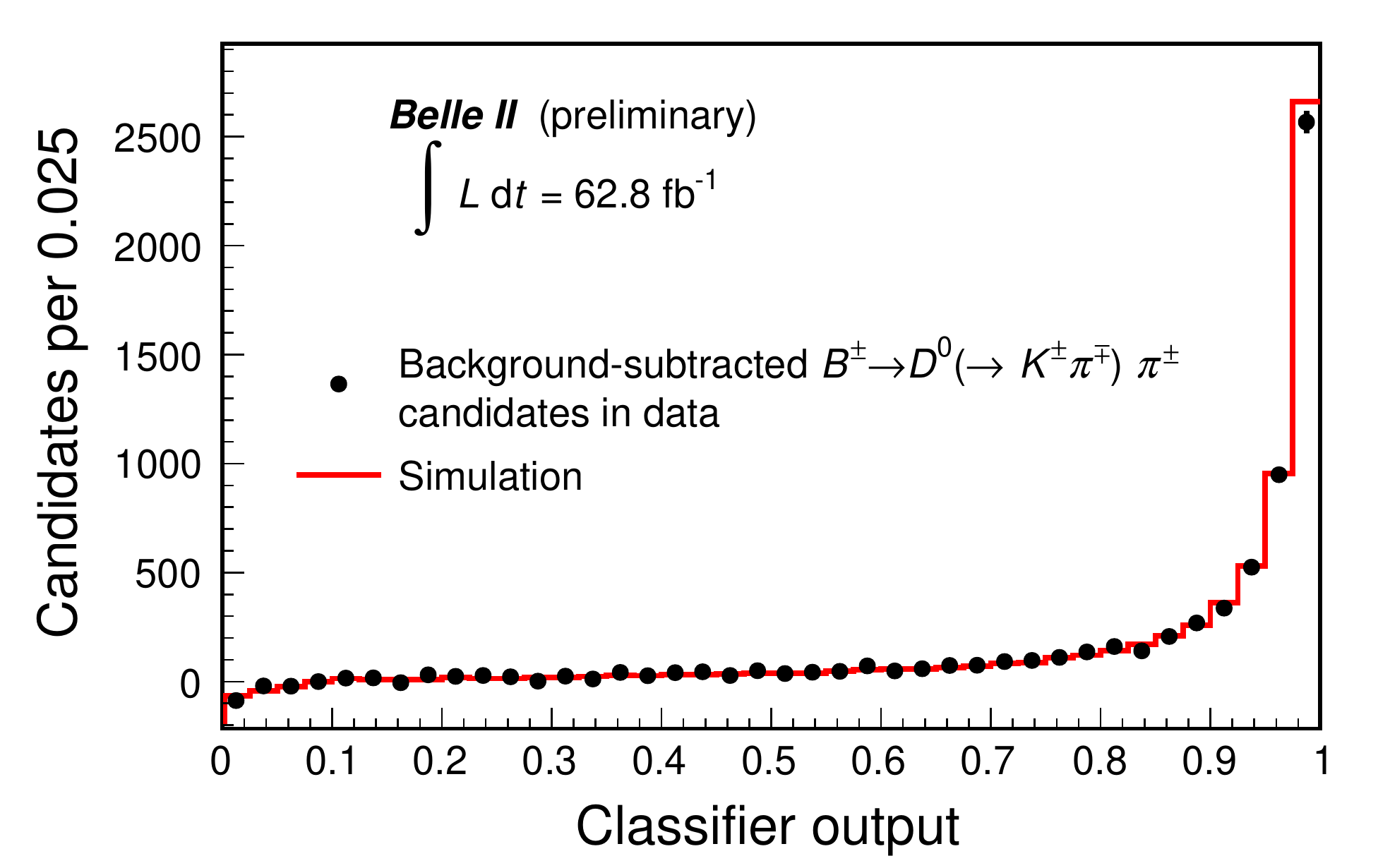}} \\
 \caption{Data-simulation comparison of the output of the boosted decision-tree classifier on (left)~sideband and (right)~sideband-subtracted $\PBplus\to\APD^{0}(\to \PKp\Pgpm)\,\Pgpp$~candidates in the signal region.}
 \label{fig:outputData_Kpi}
\end{figure}

We validate the input and output distributions of the classifier by comparing  data with simulation using control samples.
Figure~\ref{fig:outputData_Kpi} shows the distribution of the output for  \mbox{$\PBplus\to\APD^{0}(\to \PKp\Pgpm)\,\Pgpp$}~candidates reconstructed in data and simulation. No inconsistency is observed. 

\section{Optimization of the signal selection}
\label{sec:yields}
For each channel, we optimize the selection to isolate abundant, low-background signals using simulated and control-sample data. We vary the selection criteria on continuum-suppression output, charged-particle identification information, and choice of $\pi^0$ (when appropriate) to maximize ${\rm S}/\sqrt{{\rm S}+{\rm B}}$, where ${\rm S}$ and ${\rm B}$ are signal and background yields, respectively, estimated in the same signal-rich region used in the analysis. Continuum-suppression and particle-identification requirements are optimized simultaneously using simulated data. The optimal PID criteria have 77\%--86\% (channel-specific) efficiencies on kaons and 15\% misidentification rates. The $\pi^0$ selection is optimized independently by using control \mbox{$B^+ \to \overline{D}^0(\to K^+\pi^-\pi^0)\pi^+$} decays.
The optimal selection removes approximately $99\%$ of the continuum background and retains approximately $38\%$ of \mbox{$B^+ \to K^+ K^- K^+$} signal, $20\%$ of \mbox{$B^+ \to K^+ \pi^- \pi^+$} signal, and $15\%$ of \mbox{$B^0 \to K^+ \pi^- \pi^0$} signal.


\section{Determination of signal yields}
\label{sec:yields}
More than one candidate per event populates the resulting $\Delta E$ distributions, with average multiplicities up to 1.2. We restrict to one candidate per event by selecting a single $B$ candidate randomly. \par Signal yields are determined from two-dimensional maximum likelihood fits of the unbinned $\Delta E$ and $M_{\rm bc}$ distributions of candidates restricted to the signal region $M_{\rm bc} > 5.24$\,GeV/$c^2$ and $-0.15 (-0.25) < \Delta E < 0.15$ GeV in $B^+ \to h^+h^-h^+$ ($B^0 \to K^+\pi^-\pi^0$). The poorer $\Delta E$ resolution associated with the $\pi^0$ reconstruction motivates the broader $\Delta E$ range for the $B^0 \to K^+\pi^-\pi^0$ channel. Fit models are determined empirically from simulation, with additional flexibility of allowing for global shifts of peak positions and width scale-factors determined in control data, as indicated by likelihood-ratio tests. \par
We use a Gaussian function, or a sum of a Gaussian and a Crystal Ball functions~\cite{Skwarnicki:1986xj}, based on simulation to model $\Delta E$ and $M_{\rm bc}$ for all signals.  We use an exponential function and an ARGUS function~\cite{ALBRECHT1990278}, both with parameters determined in data, to model continuum background in $\Delta E$ and $M_{\rm bc}$, respectively.
We use sums of Gaussian with exponential, polynomial, or ARGUS functions, all determined from simulation, to model nonpeaking $B\bar{B}$ backgrounds. Remaining peaking backgrounds to the $B^+ \to K^+\pi^-\pi^+$ signal are modeled with Gaussian functions with shapes and normalizations constrained to the expectations from simulation. 
We model self-cross-feed~(SCF) events, in $B^0 \to K^+\pi^-\pi^0$ with the sum of a Gaussian and a Crystal Ball functions with shape and normalization constrained to the expectations from simulation. The fraction of SCF events  is around $20\%$ for $B^0 \to K^+\pi^-\pi^0$ and negligible for $B^+ \to K^+K^-K^+$ and $B^+ \to K^+\pi^-\pi^+$. \par
\begin{figure}[htb]
 \centering
 \includegraphics[width=0.475\textwidth]{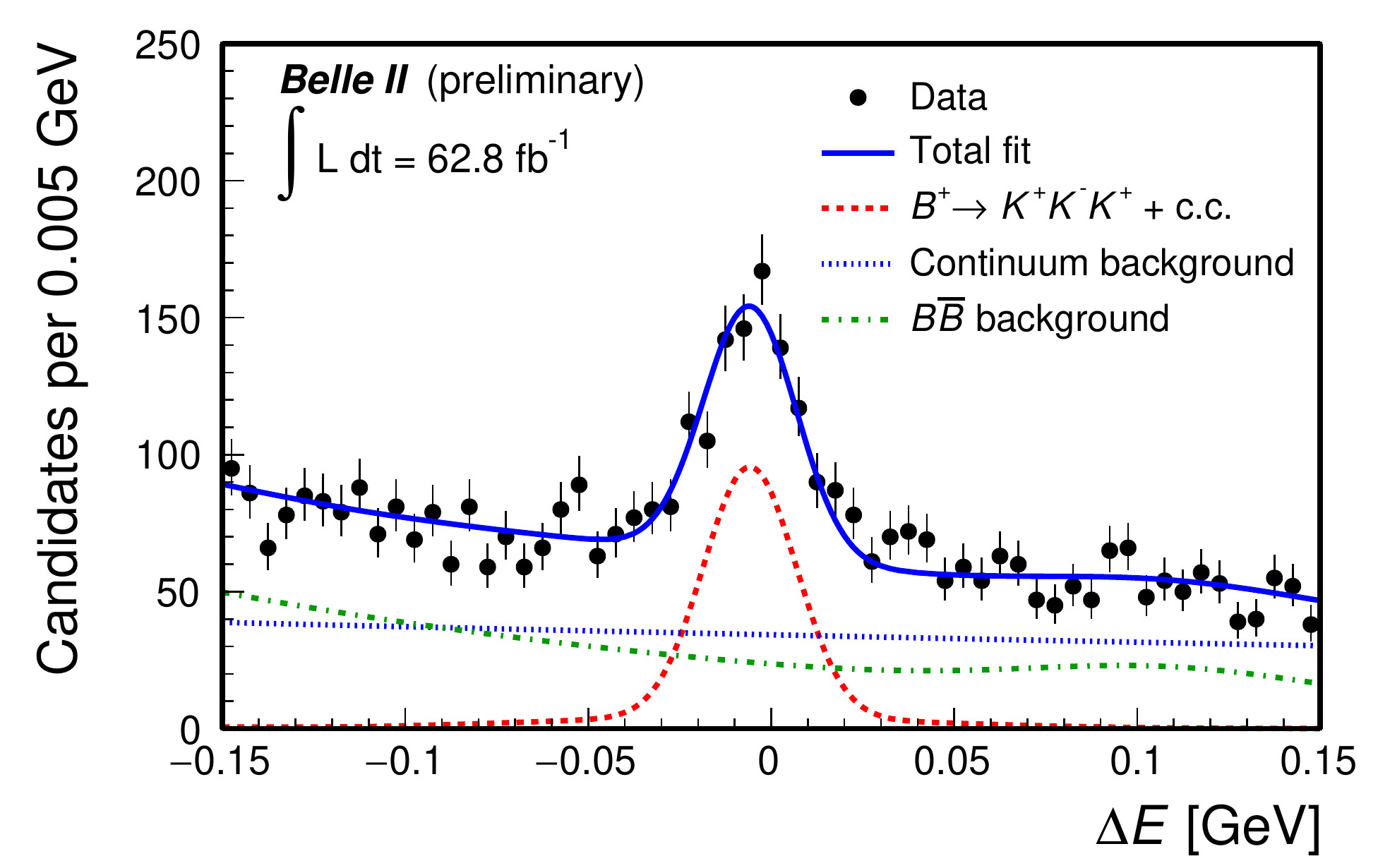}
 \includegraphics[width=0.475\textwidth]{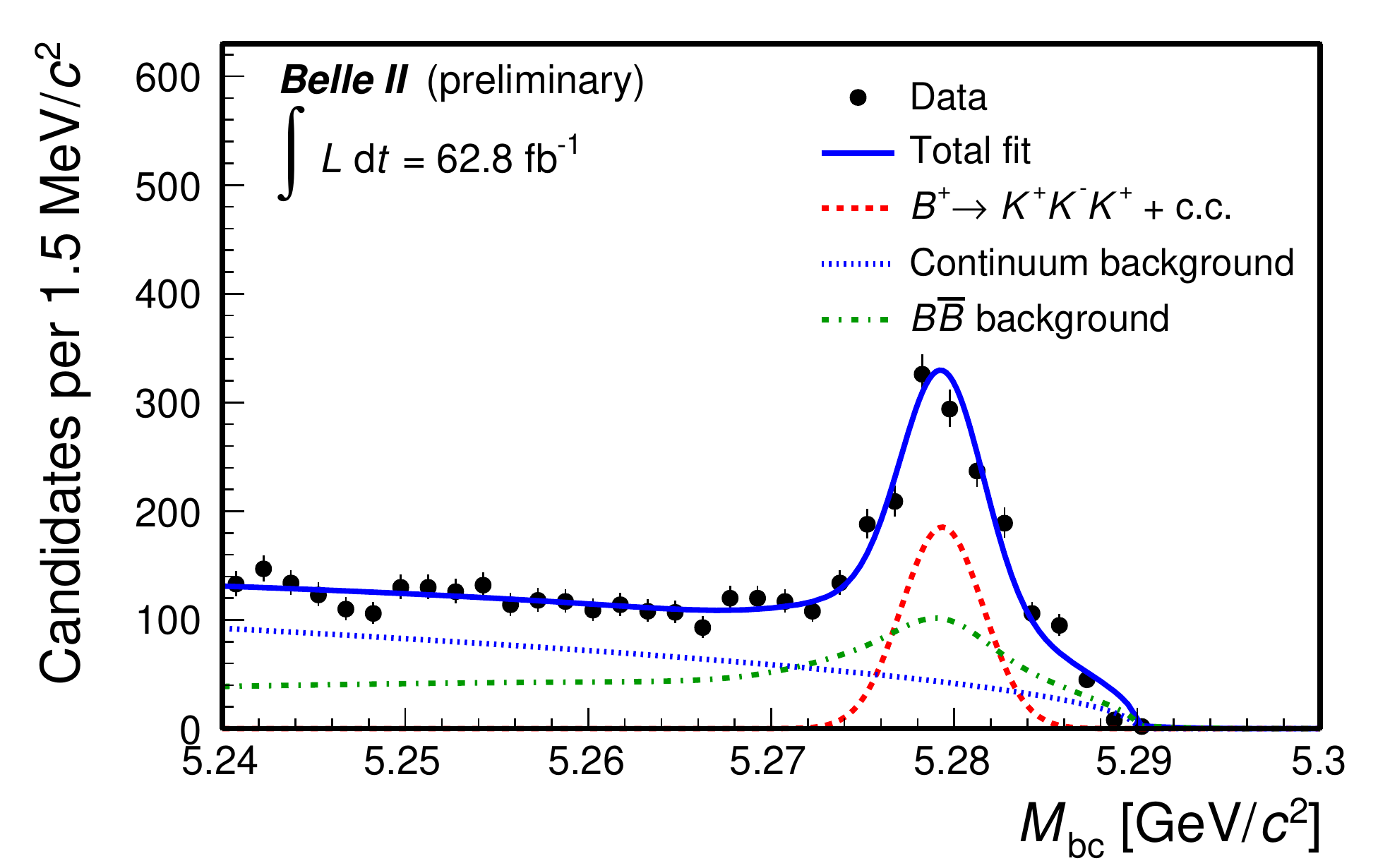}
 \caption{Distributions of (left) $\Delta E$ and (right) $M_{\rm bc}$ for $B^+ \to K^+K^-K^+$  candidates reconstructed in 2019--2020 Belle~II data, selected with an optimized continuum-suppression and kaon-enriching selection. Vetoes for peaking backgrounds are applied. Fit projections are overlaid.}
 \label{fig:KKK}
\end{figure}
\begin{figure}[htb]
 \centering
 \includegraphics[width=0.475\textwidth]{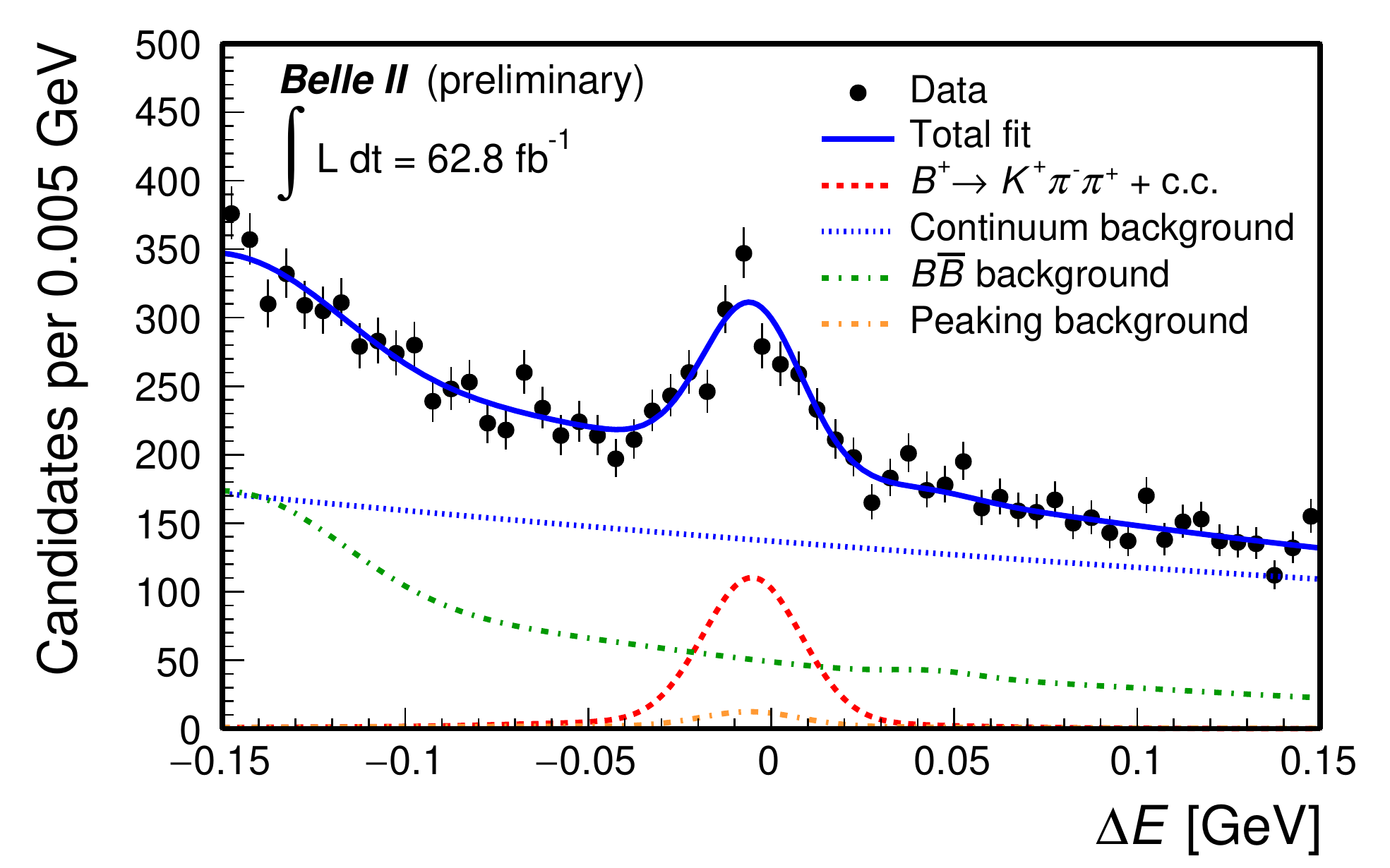}
 \includegraphics[width=0.475\textwidth]{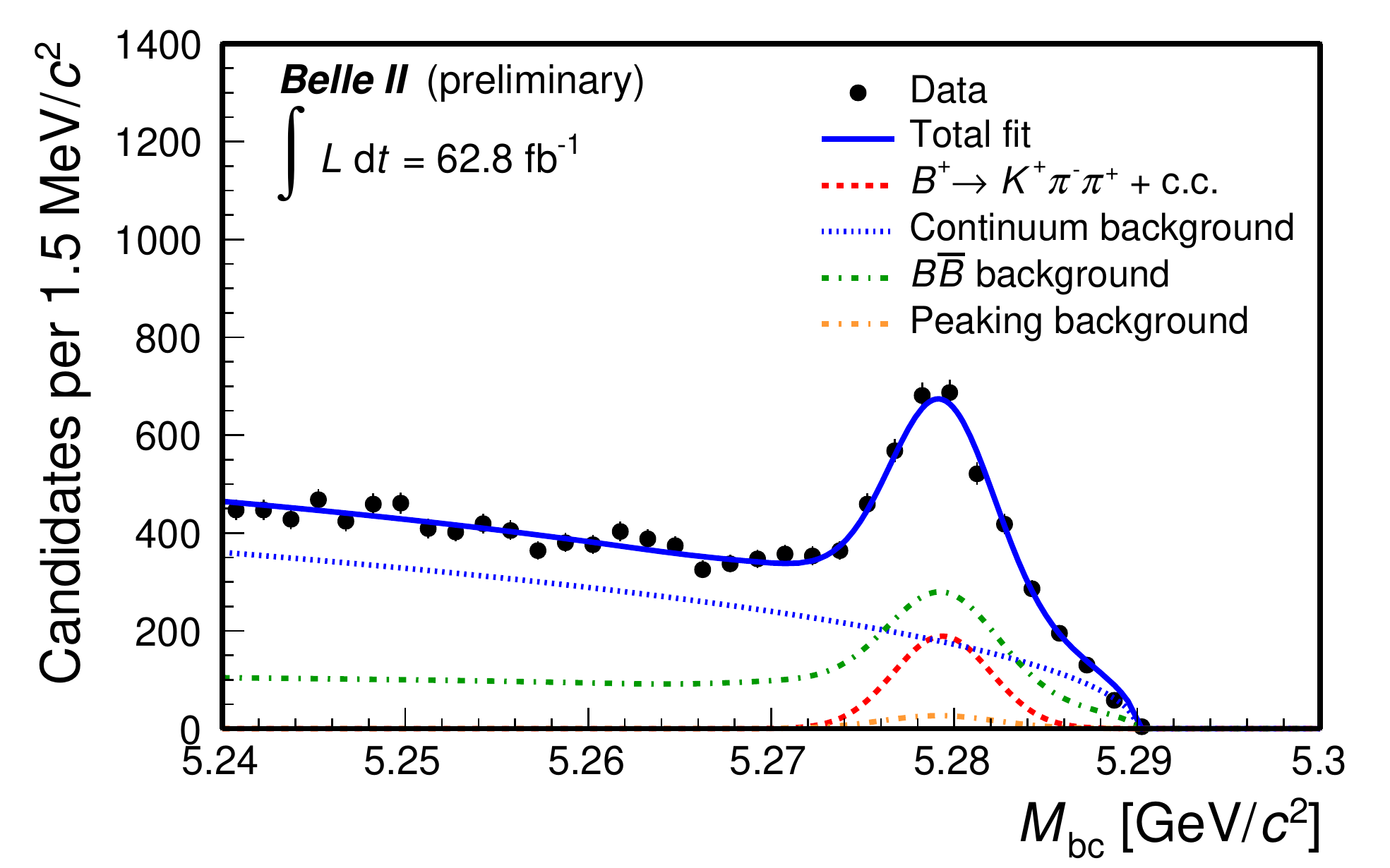}
 \caption{Distributions of (left) $\Delta E$ and (right) $M_{\rm bc}$ for $B^+ \to K^+\pi^-\pi^+$ candidates reconstructed in 2019--2020 Belle~II data, selected with an optimized continuum-suppression and kaon-enriching selection. Vetoes for peaking backgrounds are applied. Fit  projections are overlaid.}
 \label{fig:Kpipi}
\end{figure}
\begin{figure}[htb]
 \centering
 \includegraphics[width=0.475\textwidth]{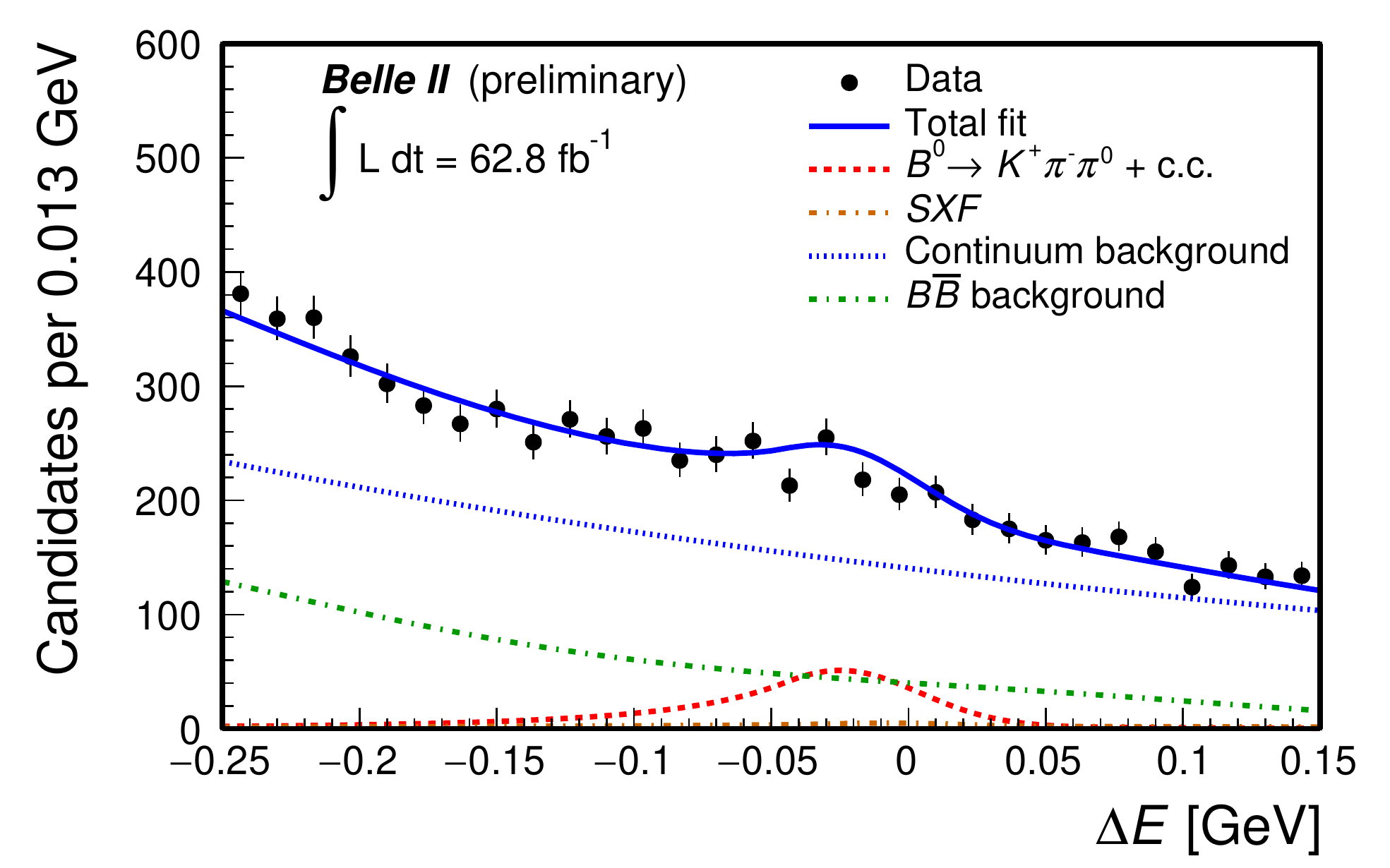}
 \includegraphics[width=0.475\textwidth]{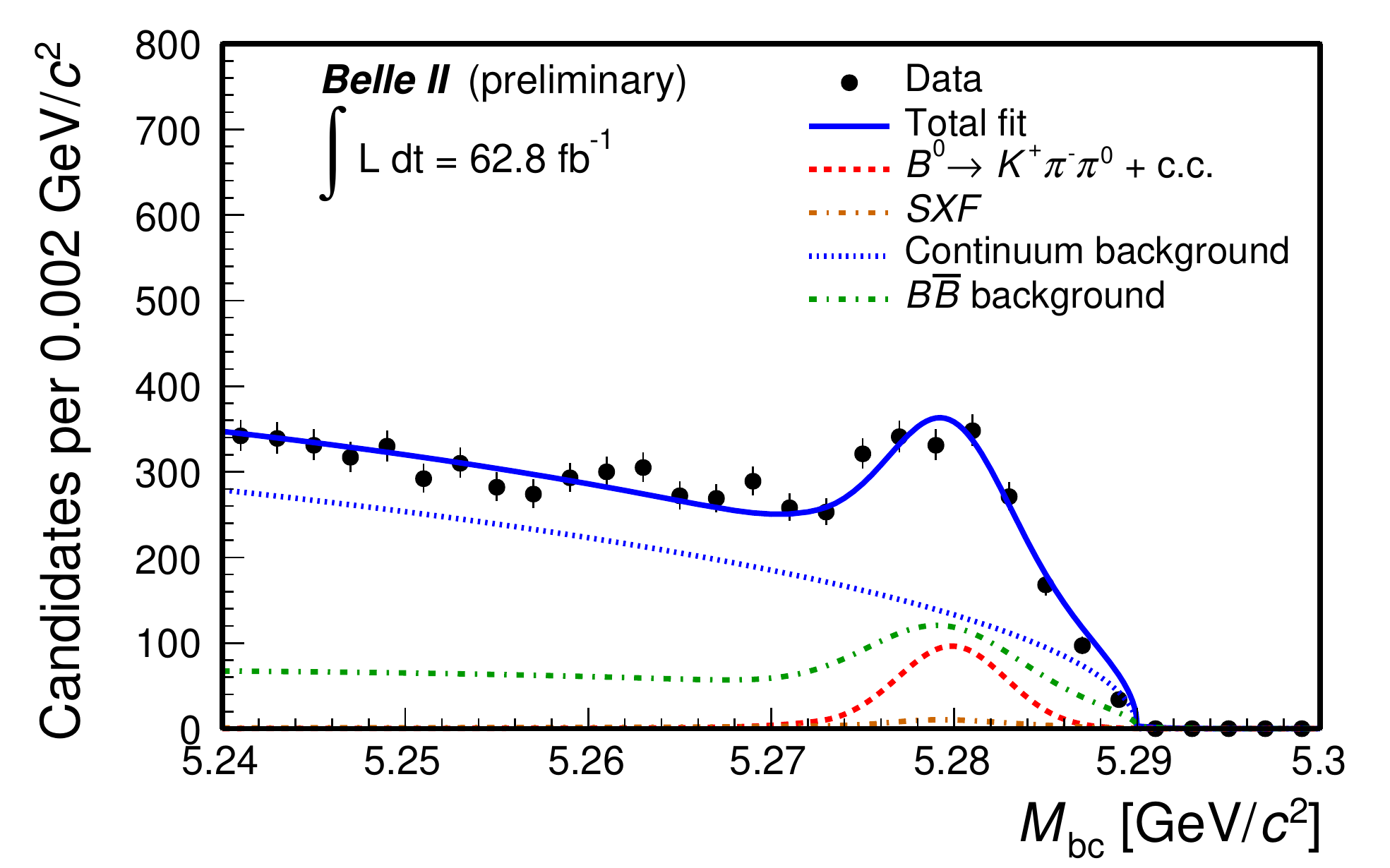}
 \caption{Distributions of (left) $\Delta E$ and (right) $M_{\rm bc}$ for $B^0 \to K^+\pi^-\pi^0$ candidates reconstructed in 2019--2020 Belle~II data, selected with an  optimized continuum-suppression and kaon-enriching selection. Vetoes for peaking backgrounds are applied. Fit  projections are overlaid.}
 \label{fig:Kpipi0}
\end{figure}
\begin{figure}[htb]
 \centering
 \includegraphics[width=0.475\textwidth]{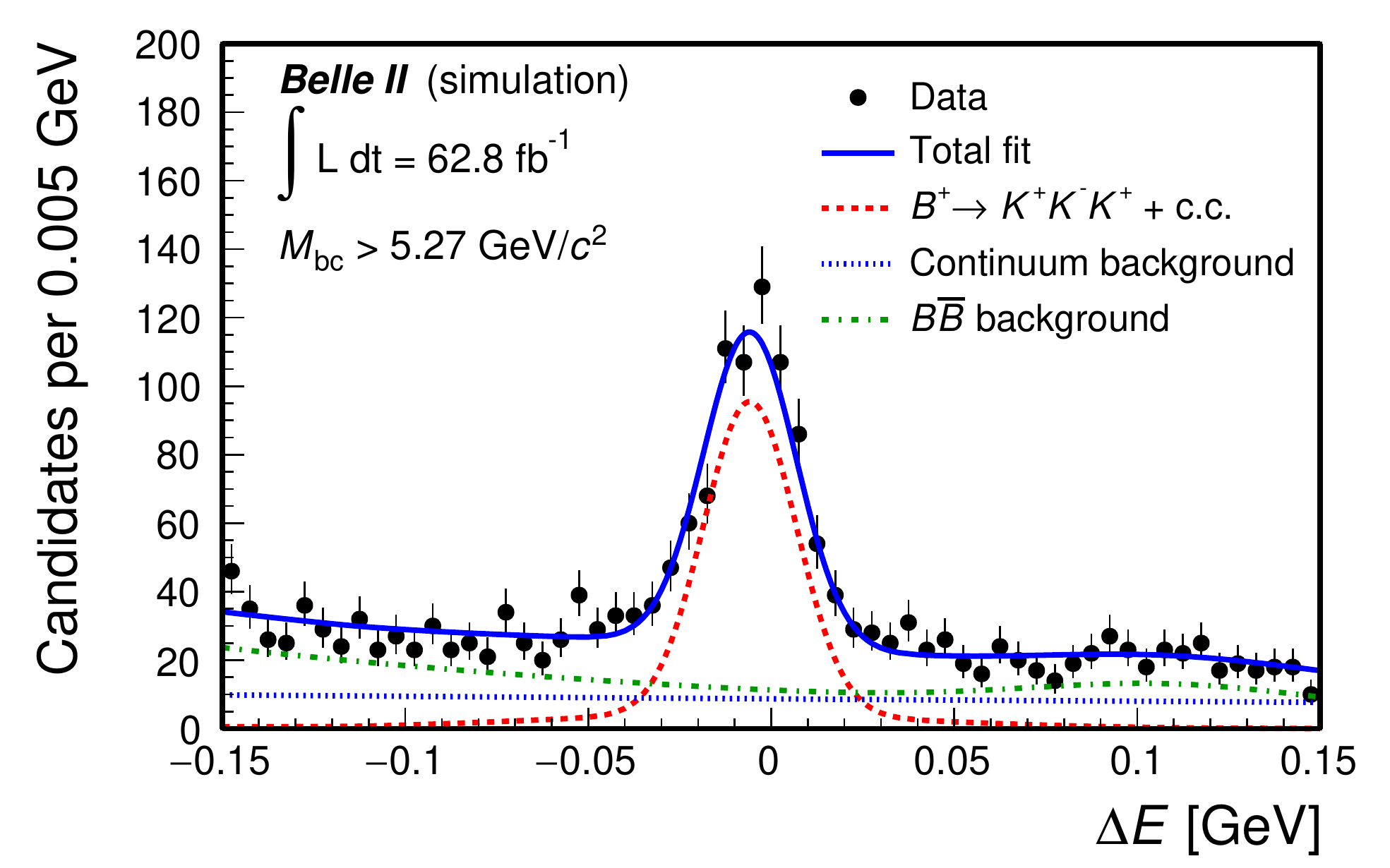}
 \includegraphics[width=0.475\textwidth]{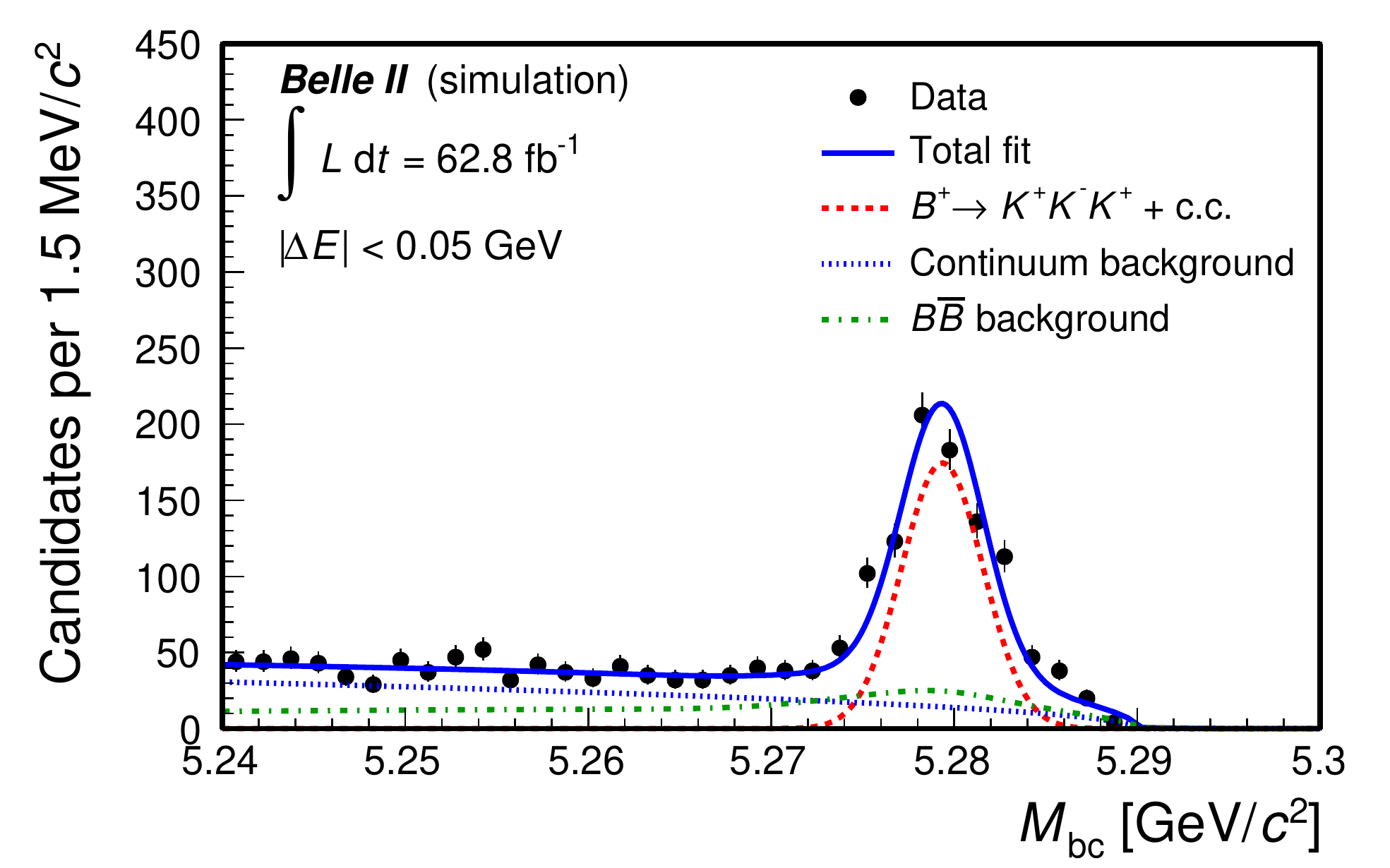}
 \caption{Signal-enhanced distributions of (left) $\Delta E$ (with $M_{\rm bc}>5.27$ GeV/$c^2$) and (right) $M_{\rm bc}$ (with $|\Delta E|<0.05$ GeV) for $B^+ \to K^+K^-K^+$  candidates reconstructed in 2019--2020 Belle~II data, selected with an optimized continuum-suppression and kaon-enriching selection. Vetoes for peaking backgrounds are applied. Fit projections are overlaid.}
 \label{fig:KKK_SigE}
\end{figure}
\begin{figure}[htb]
 \centering
 \includegraphics[width=0.475\textwidth]{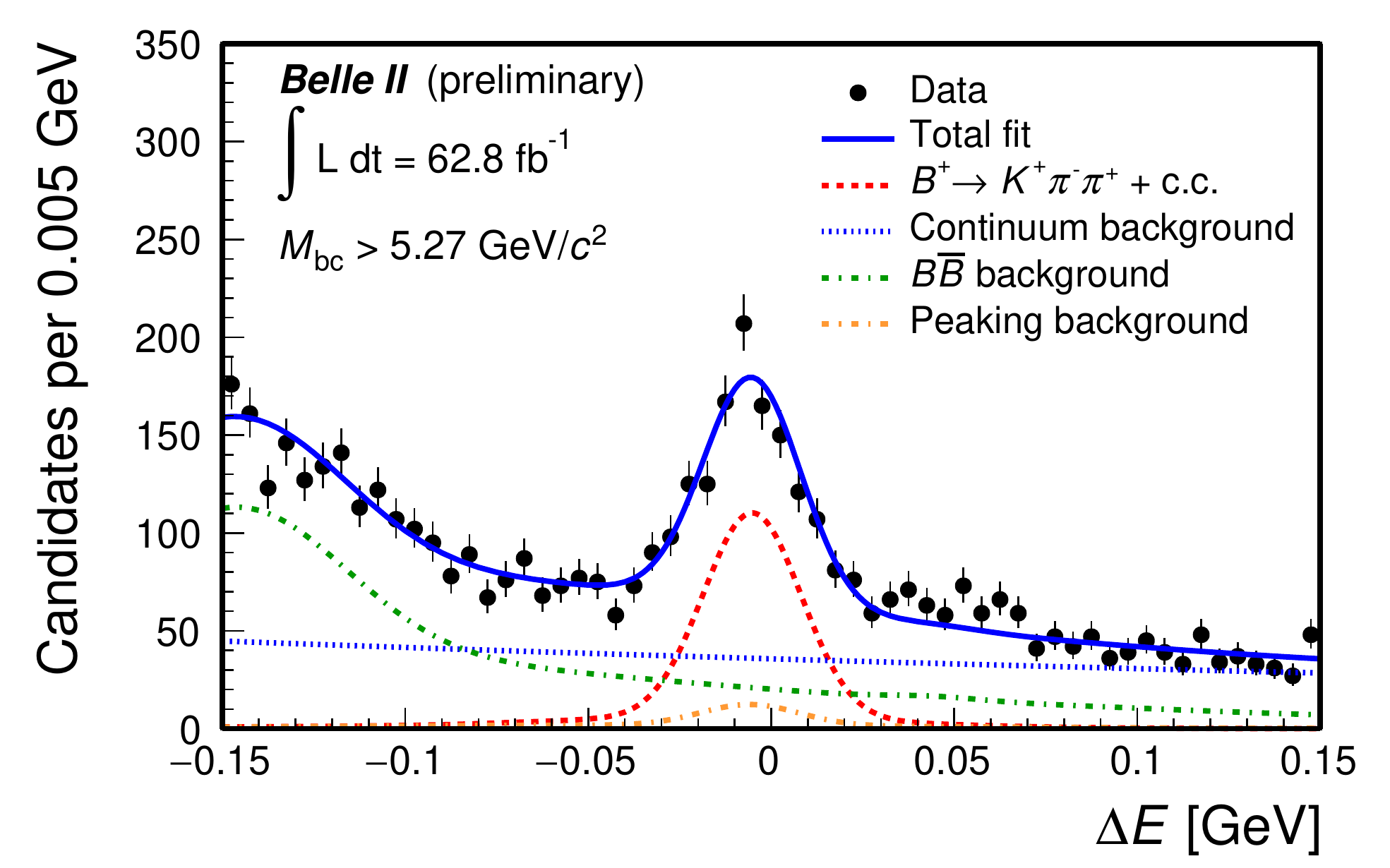}
 \includegraphics[width=0.475\textwidth]{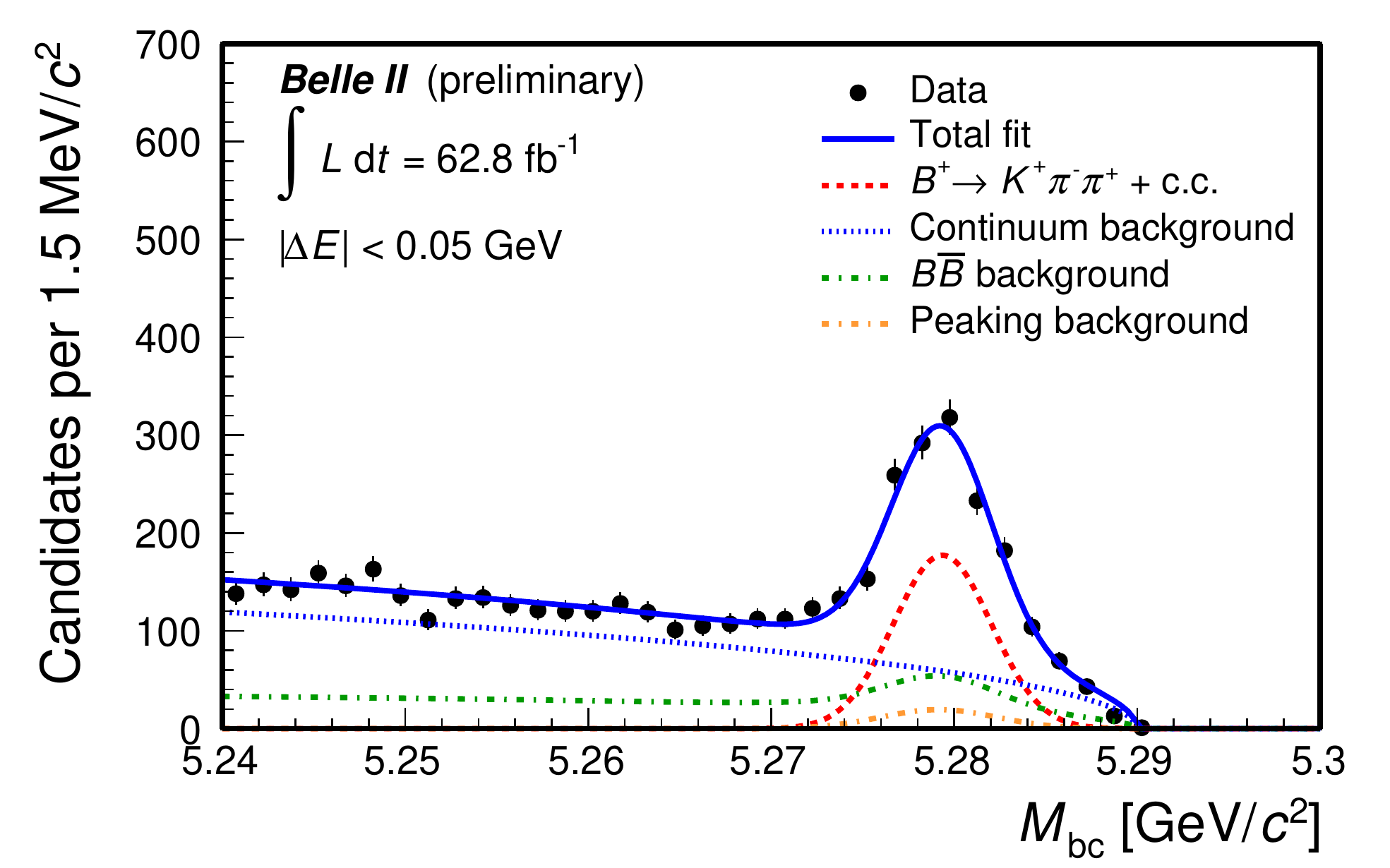}
 \caption{Signal-enhanced distributions of (left) $\Delta E$ (with $M_{\rm bc}>5.27$ GeV/$c^2$) and (right) $M_{\rm bc}$ (with $|\Delta E|<0.05$ GeV) for $B^+ \to K^+\pi^-\pi^+$ candidates reconstructed in 2019--2020 Belle~II data, selected with an  optimized continuum-suppression and kaon-enriching selection. Vetoes for peaking backgrounds are applied. Fit projections are overlaid.}
 \label{fig:Kpipi_SigE}
\end{figure}
\begin{figure}[htb]
 \centering
 \includegraphics[width=0.475\textwidth]{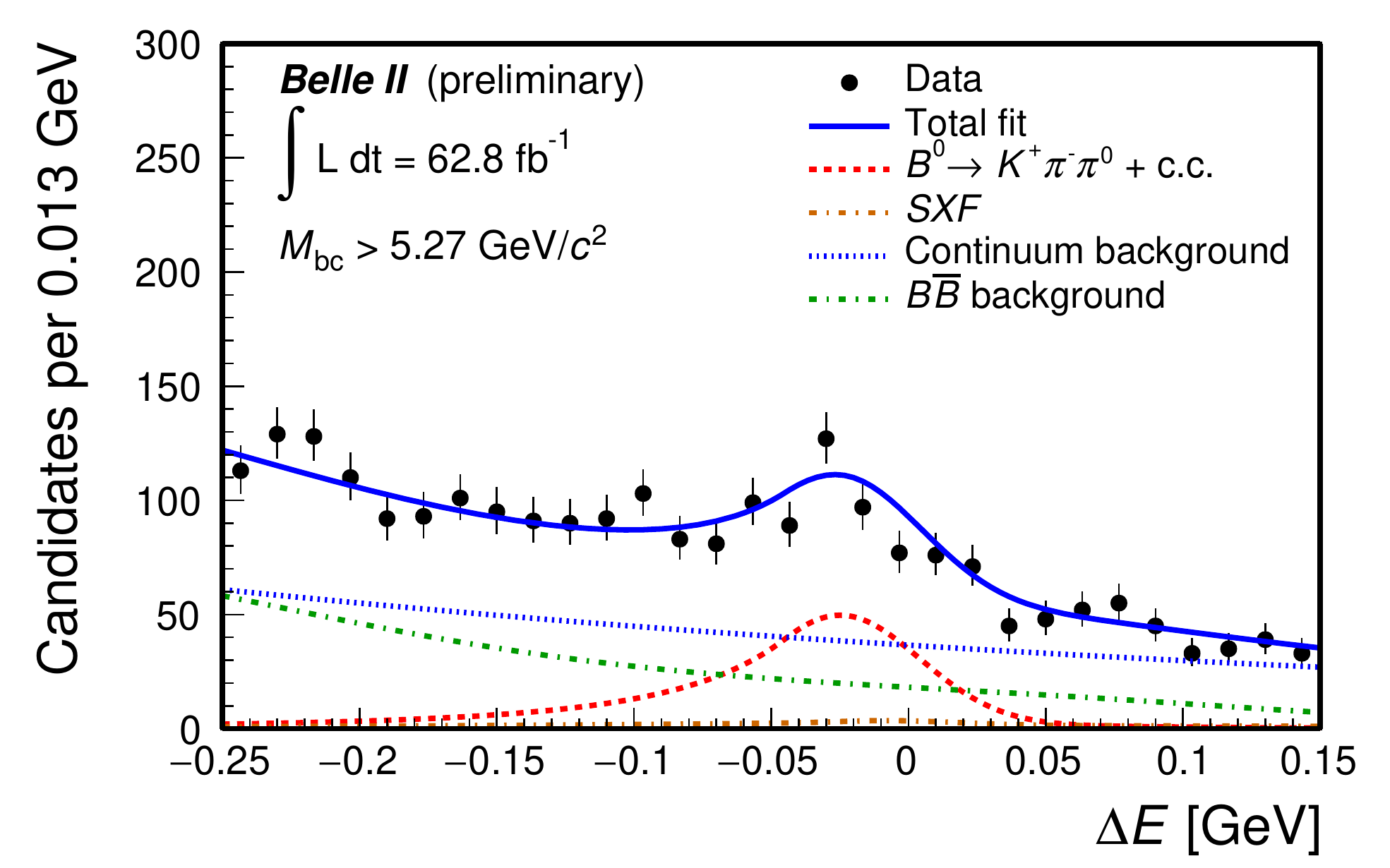}
 \includegraphics[width=0.475\textwidth]{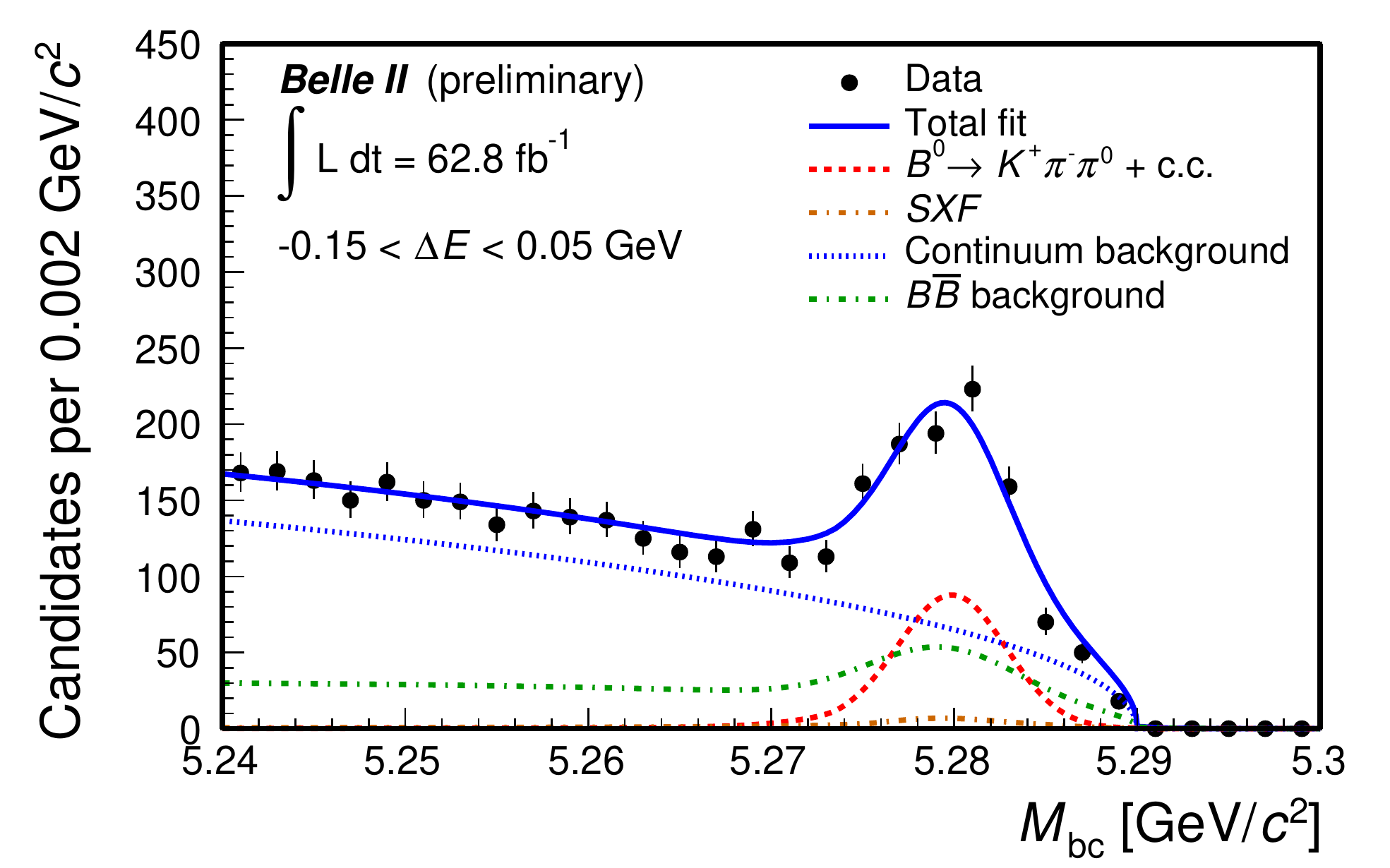}
 \caption{Signal-enhanced distributions of (left) $\Delta E$ (with $M_{\rm bc}>5.27$ GeV/$c^2$) and (right) $M_{\rm bc}$ (with $-0.15<\Delta E<0.05$ GeV) for $B^0 \to K^+\pi^-\pi^0$ candidates reconstructed in 2019--2020 Belle~II data, selected with an  optimized continuum-suppression and kaon-enriching selection. Vetoes for peaking backgrounds are applied. Fit projections are overlaid.}
 \label{fig:Kpipi0_SigE}
\end{figure}

The $\Delta E$ and $M_{\rm bc}$ distributions with fit projections overlaid are shown in Figs.~\ref{fig:KKK}--\ref{fig:Kpipi0}. Figures~\ref{fig:KKK_SigE}--\ref{fig:Kpipi0_SigE} show the corresponding signal-enhanced distributions.
Narrow peaking signals are visible overlapping smooth backgrounds, mostly dominated by continuum. The $\Delta E$ distribution of $B^0 \to K^+ \pi^- \pi^0$ candidates has a low-$\Delta E$ tail, presumably due to energy leakage from calorimeter crystals, which affects  $\pi^0$ reconstruction. 


\begin{figure}[htb]
 \centering
 \includegraphics[width=0.475\textwidth]{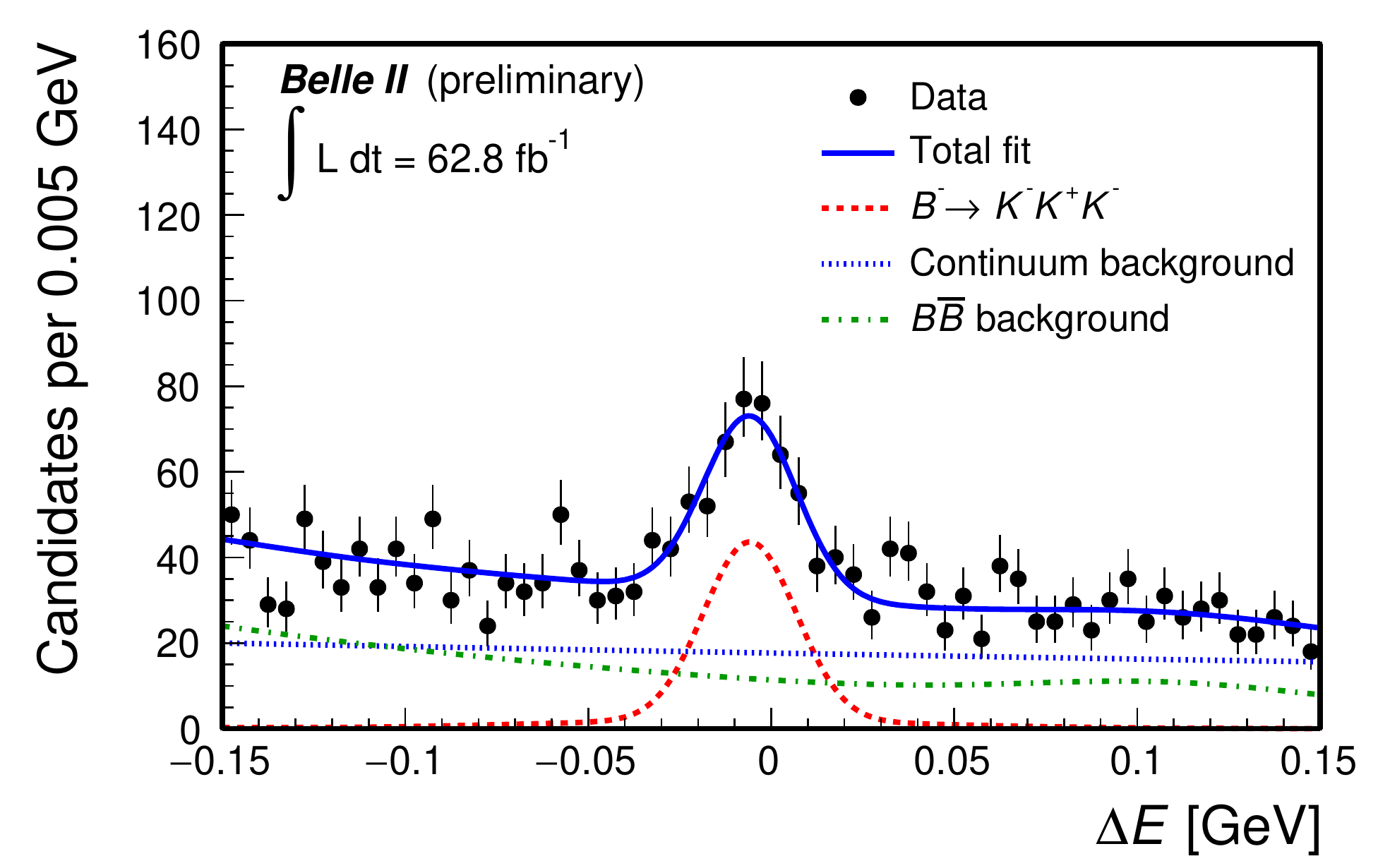}
 \includegraphics[width=0.475\textwidth]{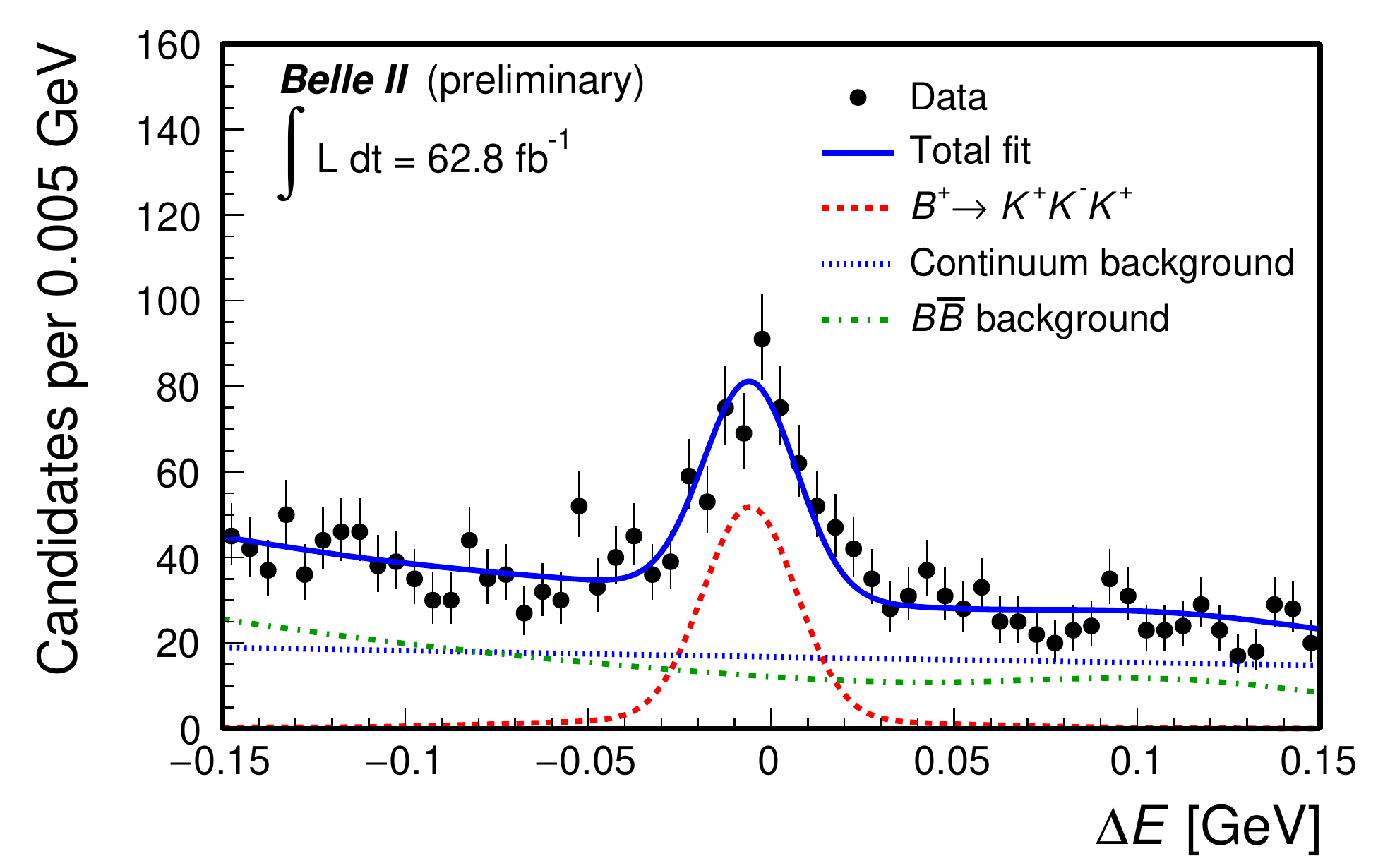}
 \includegraphics[width=0.475\textwidth]{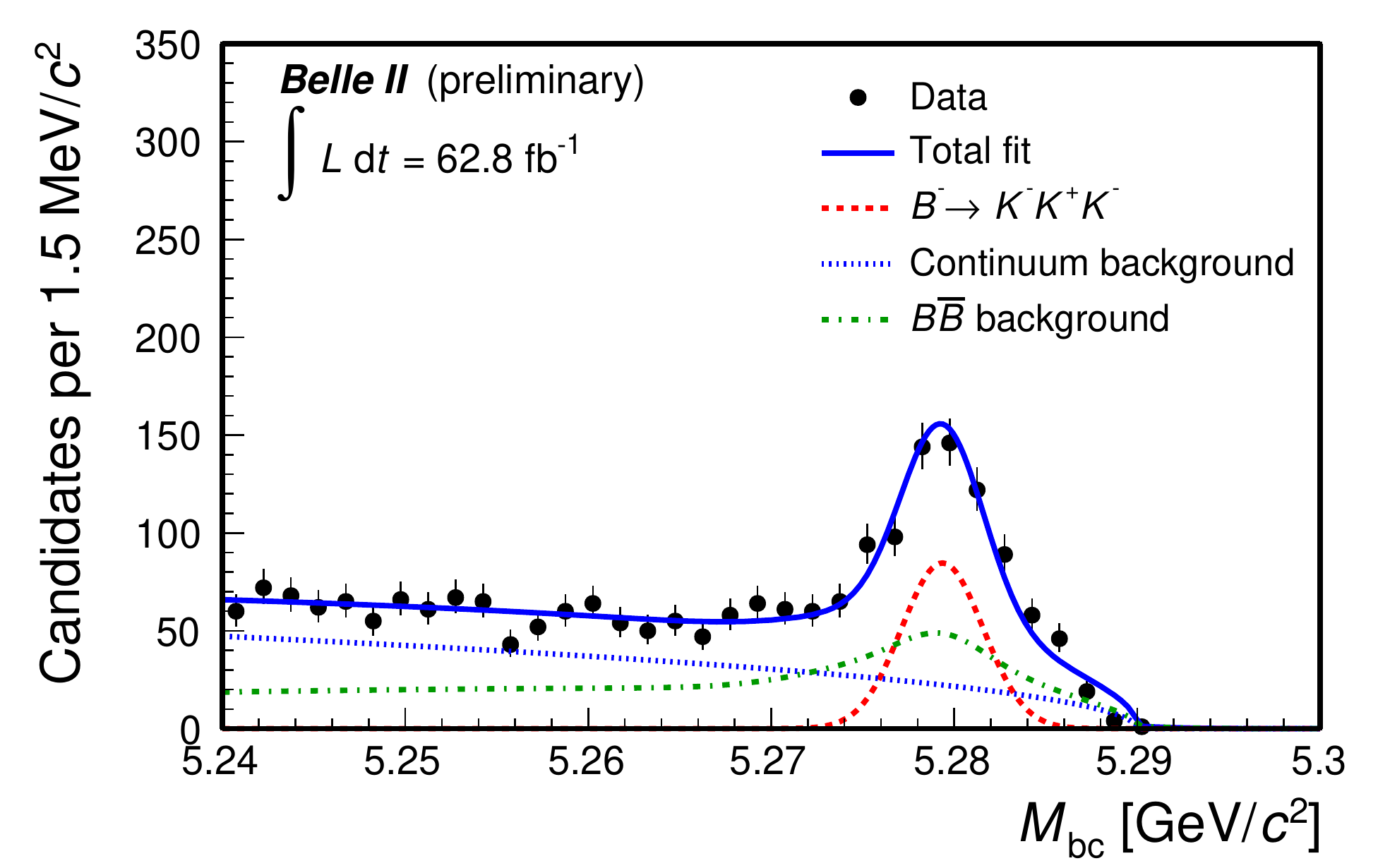}
 \includegraphics[width=0.475\textwidth]{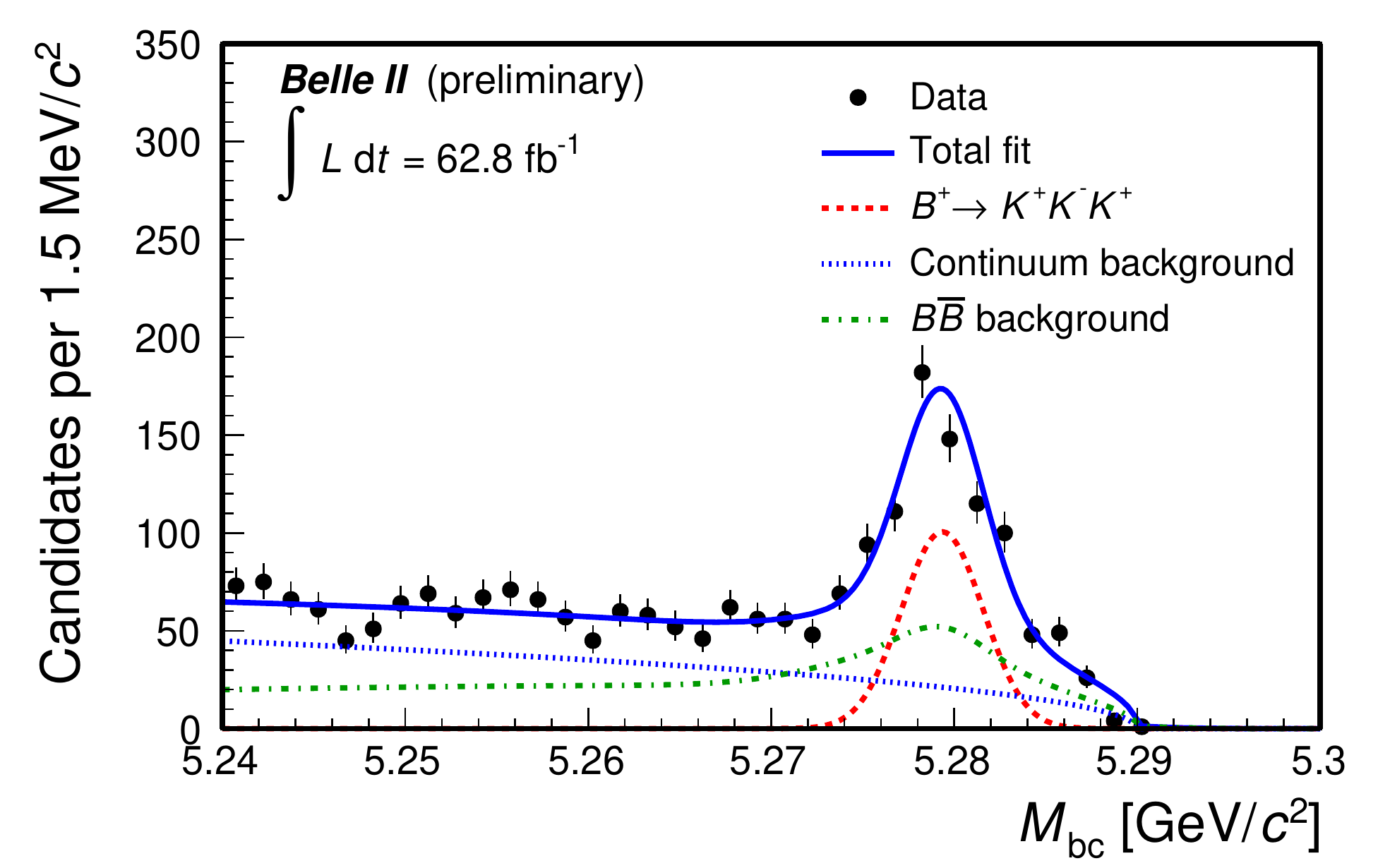}
 \caption{Distributions of (top) $\Delta E$ and (bottom) $M_{\rm bc}$ for (left) $B^+ \to K^+K^-K^+$ and (right) $B^- \to K^- K^+K^-$ candidates reconstructed in 2019--2020 Belle~II data selected with an  optimized continuum-suppression and kaon-enriching selection. Vetoes for peaking backgrounds are applied.  Fit projections are overlaid.}
 \label{fig:ACP_KKK}
\end{figure}

\begin{figure}[htb]
 \centering
 \includegraphics[width=0.475\textwidth]{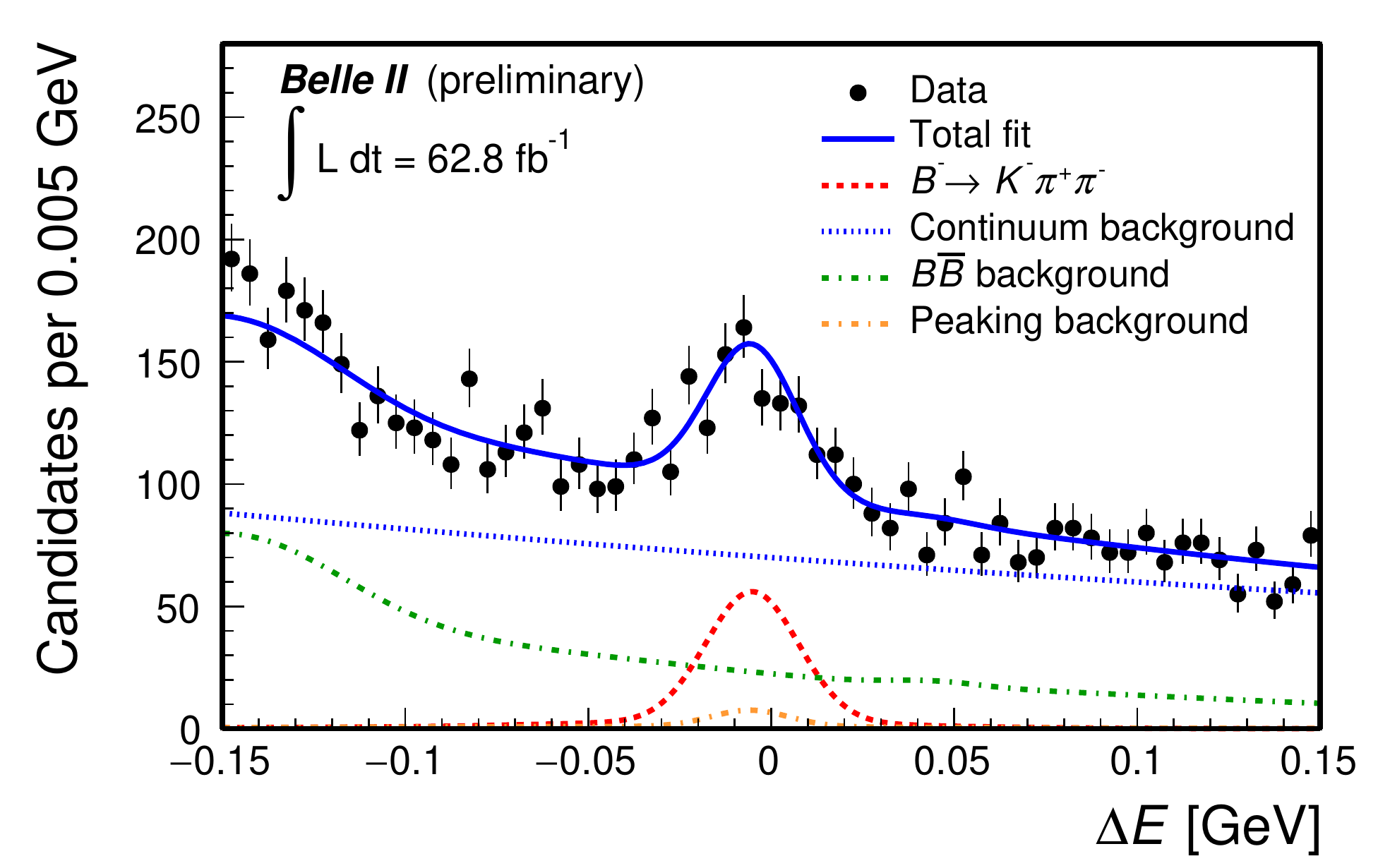}
 \includegraphics[width=0.475\textwidth]{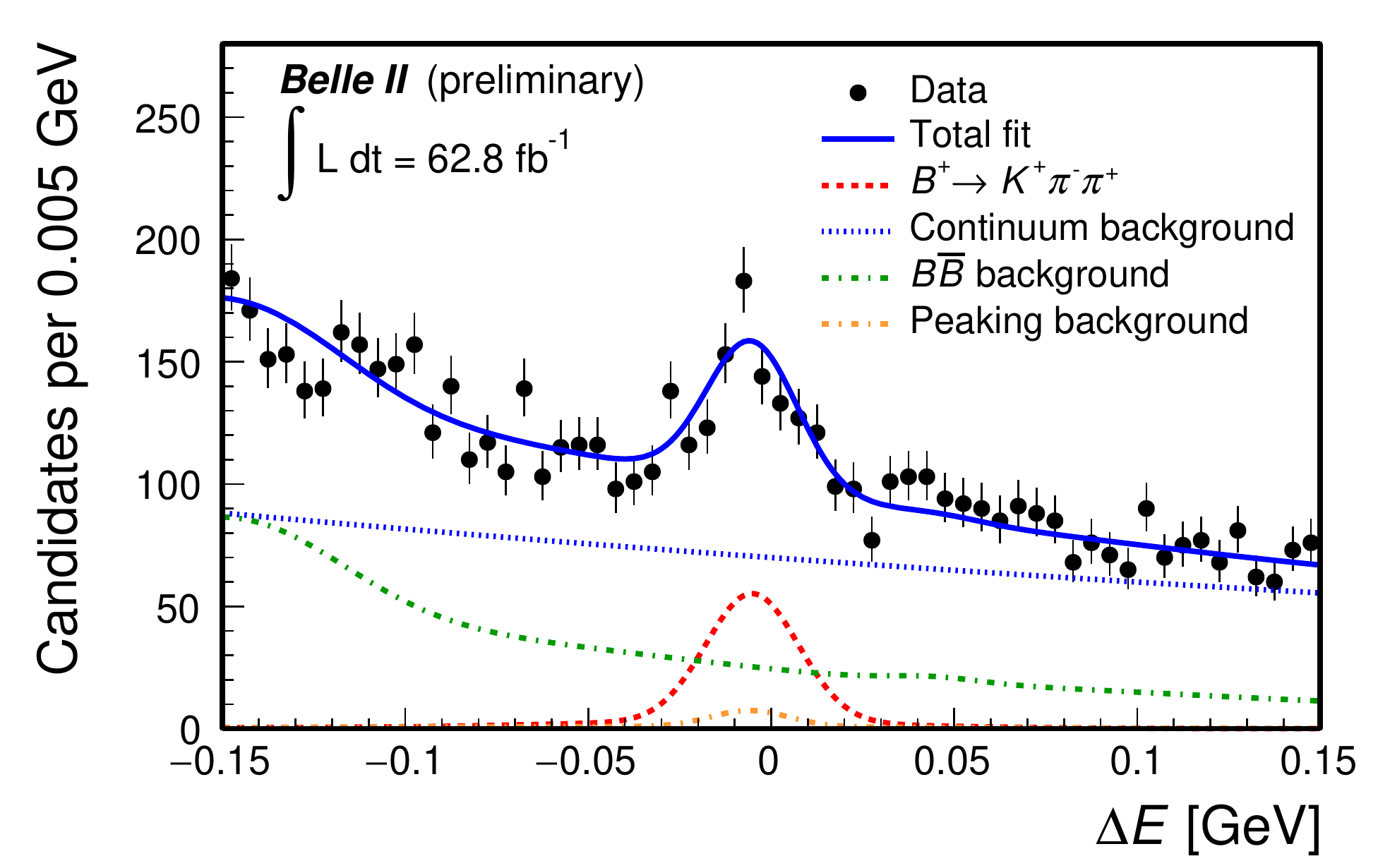}
 \includegraphics[width=0.475\textwidth]{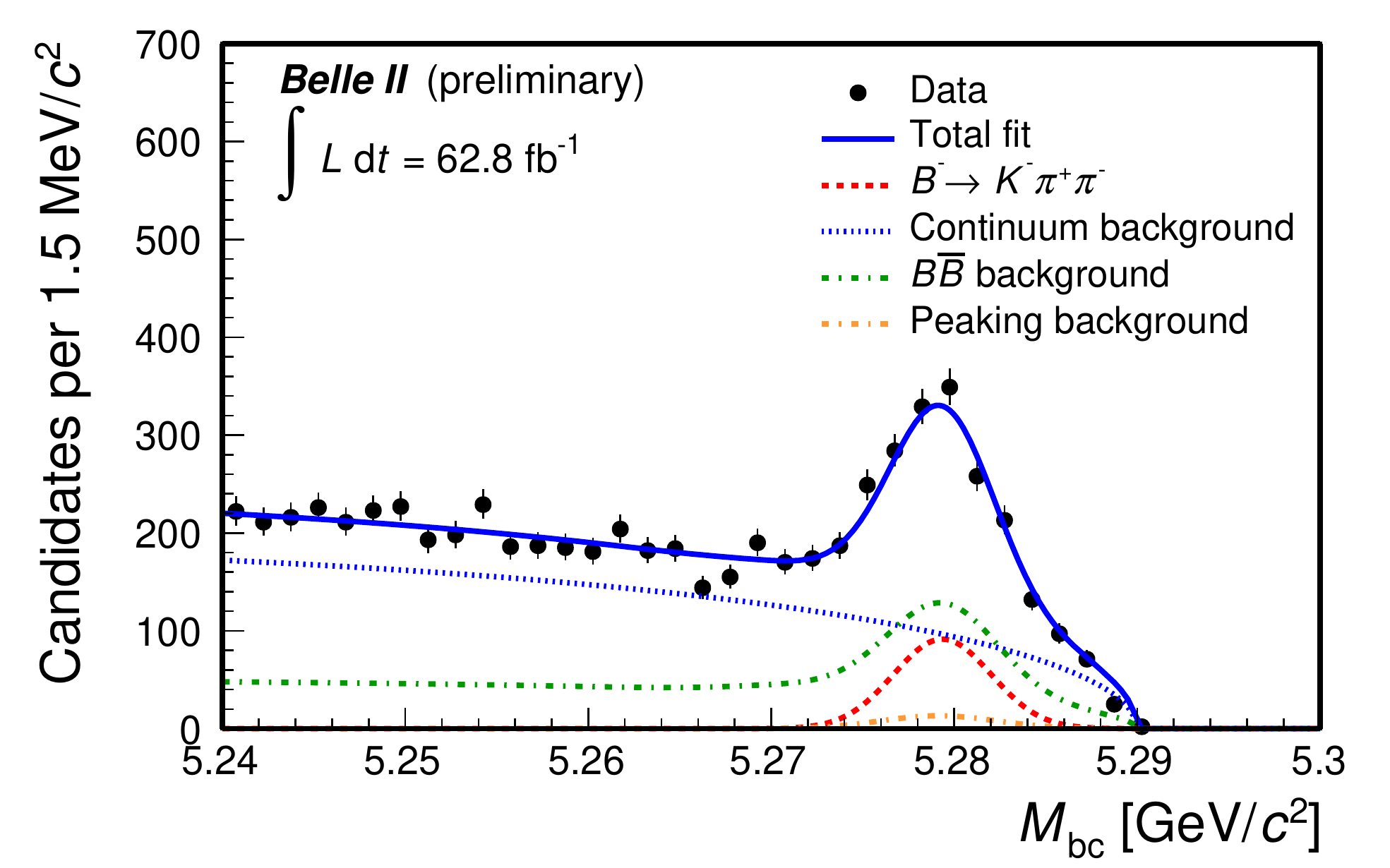}
 \includegraphics[width=0.475\textwidth]{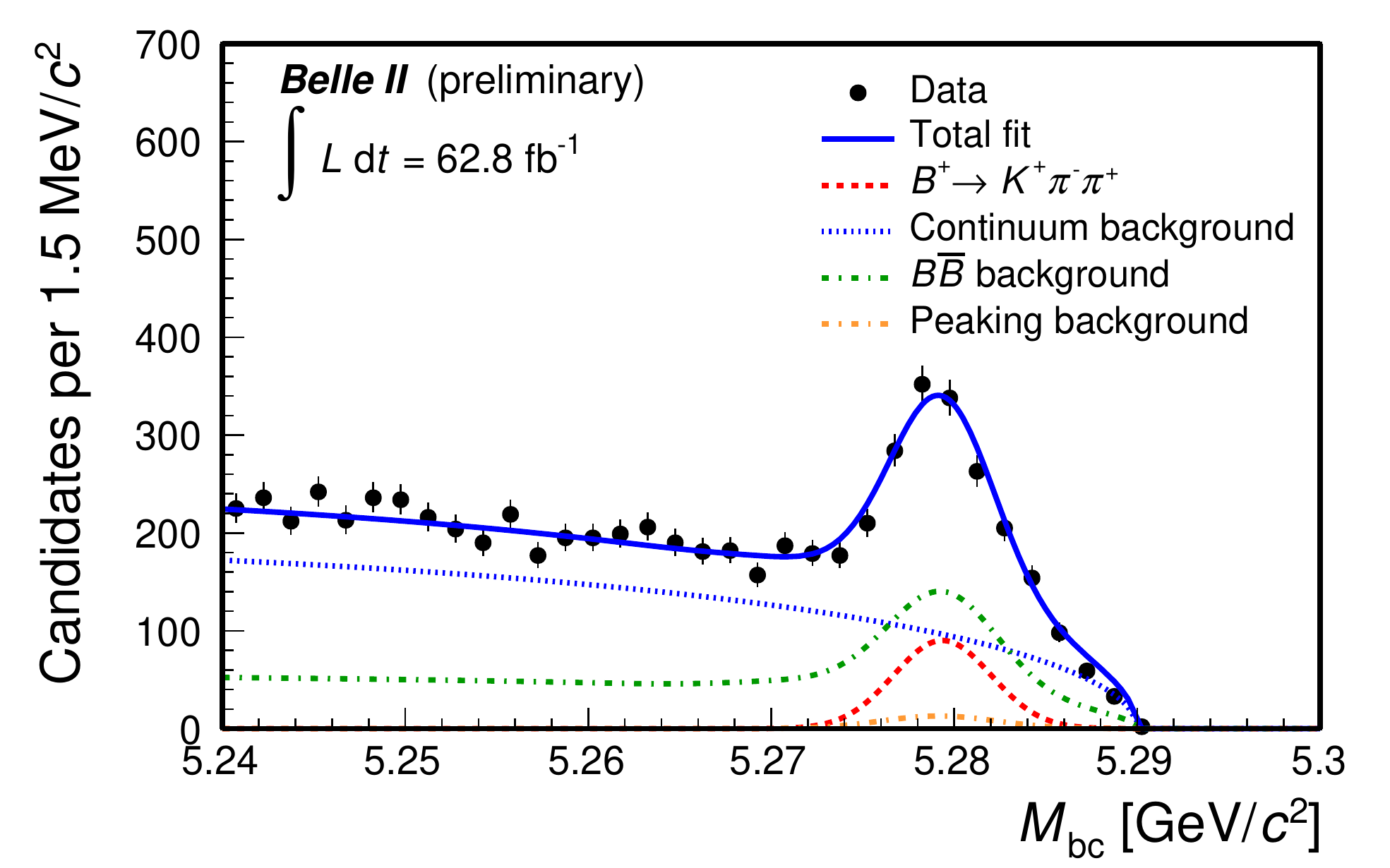}
 \caption{Distributions of (top) $\Delta E$ and (bottom) $M_{\rm bc}$ for (left) $B^+ \to K^+\pi^-\pi^+$ and (right) $B^- \to K^- \pi^+\pi^-$ candidates reconstructed in 2019--2020 Belle~II data selected with an optimized continuum-suppression and kaon-enriching selection. Vetoes for peaking backgrounds are applied. Fit  projections are overlaid.}
 \label{fig:ACP_Kpipi}
\end{figure}

\begin{figure}[htb]
 \centering
 \includegraphics[width=0.475\textwidth]{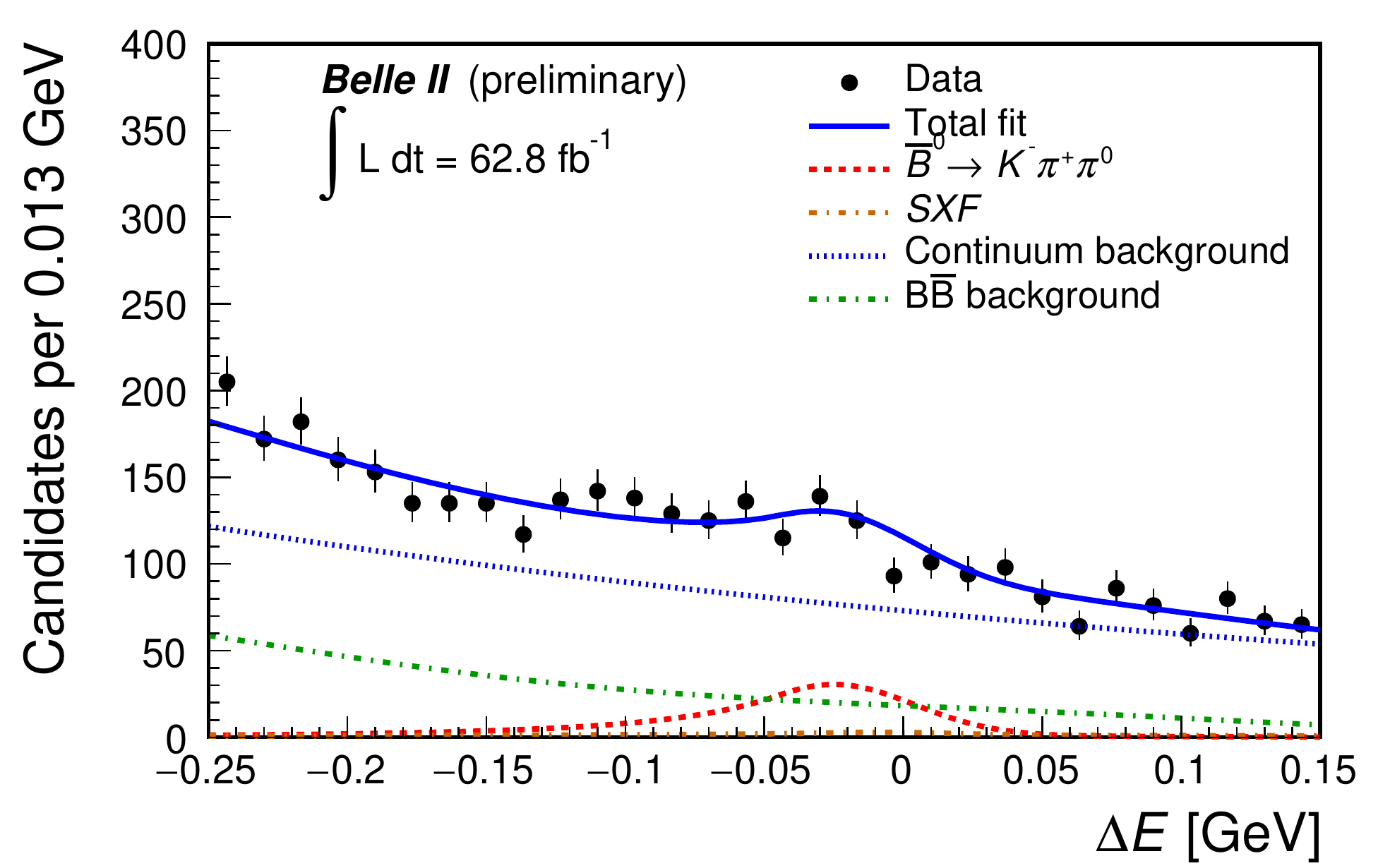}
 \includegraphics[width=0.475\textwidth]{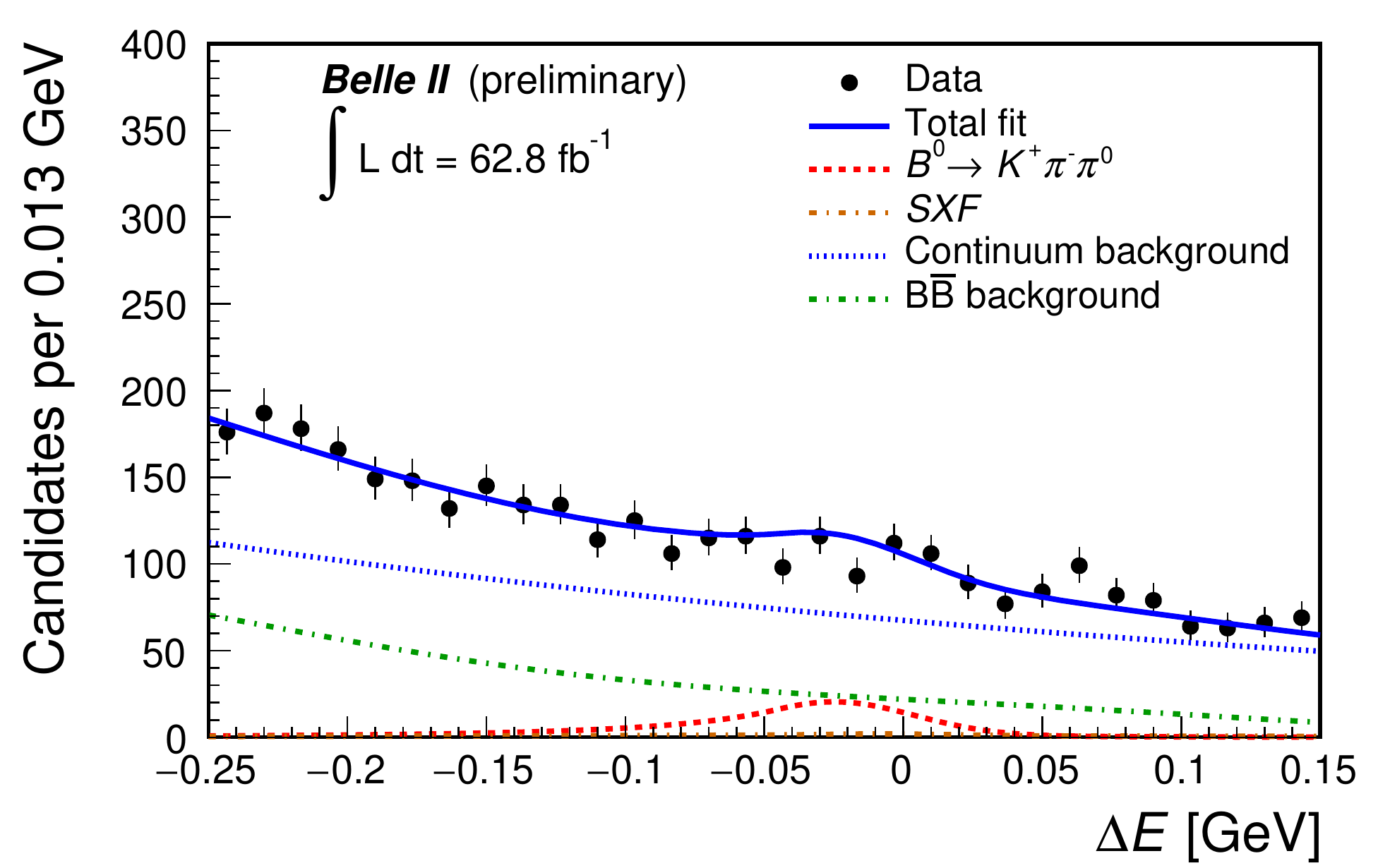}
  \includegraphics[width=0.475\textwidth]{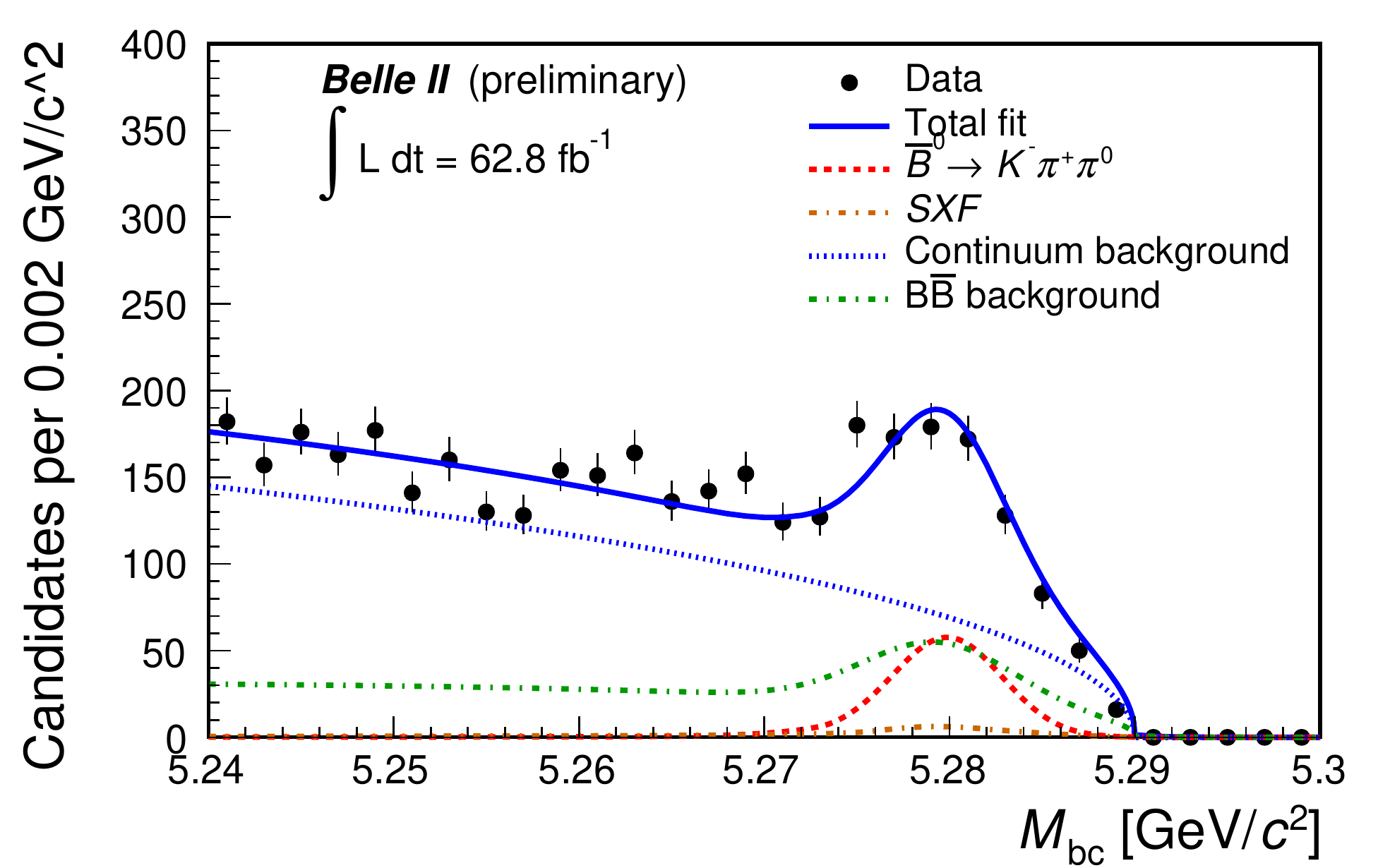}
 \includegraphics[width=0.475\textwidth]{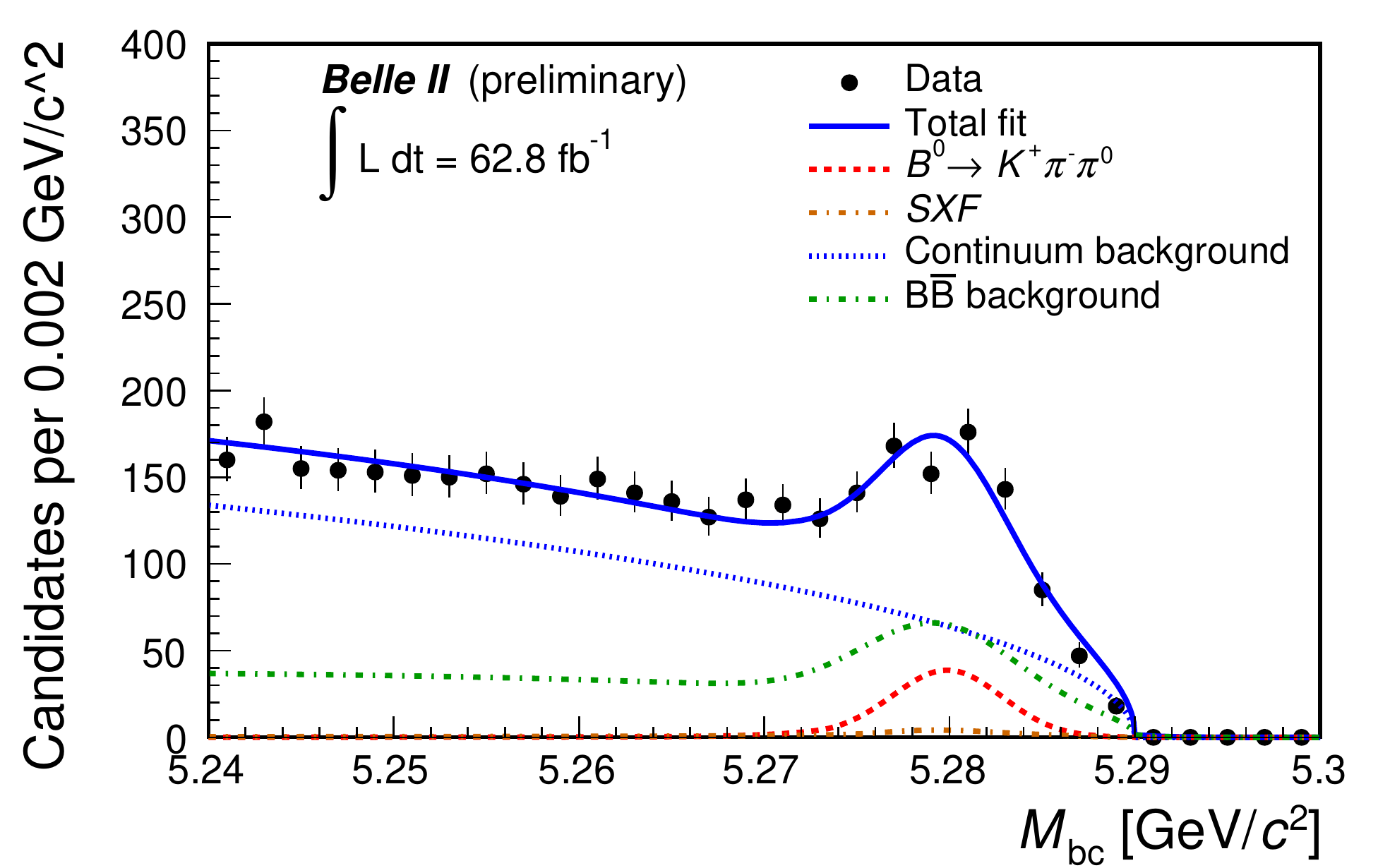}
 \caption{Distributions of (top) $\Delta E$ and (bottom) $M_{\rm bc}$ for (left) $\overline{B}^0 \to K^- \pi^+\pi^0$ and (right) $B^0 \to K^+\pi^-\pi^0$ candidates reconstructed in 2019--2020 Belle~II data selected with an optimized continuum-suppression and kaon-enriching selection. Vetoes for peaking backgrounds are applied. Fit projections are overlaid.}
 \label{fig:ACP_Kpipi0}
\end{figure}

\clearpage
In addition, we use a nonextended likelihood to fit simultaneously the unbinned $\Delta E$ and $M_{\rm bc}$ distributions of bottom and antibottom candidates decaying in flavor-specific final states for measurements of direct CP violation. We use the same signal and background models as for branching-fraction measurements and determine directly the raw charge-dependent yield asymmetry as a fit parameter,
\begin{equation*}
    \mathcal{A}=\frac{N(b)-N(\bar{b})}{N(b)+N(\bar{b})},
\end{equation*}
where $N$ are signal yields and $b$ ($\bar{b}$) indicates the meson containing a bottom (antibottom) quark. 
Charge-specific $\Delta E$ and $M_{\rm bc}$ distributions are shown in Figs.~\ref{fig:ACP_KKK}--\ref{fig:ACP_Kpipi0} with fit projections overlaid.

 \begin{table}[!ht]
 \caption{Summary of charge-specific signal yields for the measurement of CP-violating asymmetries in 2019-2020 Belle II data. Only statistical uncertainties are reported.} 
     \label{tab:ACP}
     \centering
 \begin{tabular}{l  r  r  r }
 \hline\hline
 \multicolumn{1}{c}{} & 
 \multicolumn{2}{c}{Yield} &   
 \multicolumn{1}{c}{Raw asymmetry} \\
 Decay & \multicolumn{1}{c}{$B^+$} & \multicolumn{1}{c}{$B^-$} & \multicolumn{1}{c}{}  \\\hline
   $B^+ \to K^+ K^- K^+$    &	$375 \pm 22$ &	$315\pm 20$ &	$-0.086\pm 0.042$ \\ 
   $B^+ \to K^+ \pi^- \pi^+$    &	$419 \pm 30$ &	$424\pm 30$ &	$0.007\pm 0.050$  \\
   $B^0 \to K^+ \pi^- \pi^0$    &	$152 \pm 23$ &	$227\pm 25$ &	$0.197\pm 0.088$  \\
   \hline
 \end{tabular}
\end{table}{}

\clearpage
\section{Efficiencies and corrections}
The raw event yields observed in data are corrected for selection and reconstruction effects to obtain physics quantities. 
We determine the signal efficiency for each channel as the sum of the efficiencies for all major submodes contributing to the Dalitz plot, as determined by Belle II simulation, weighted for submode abundances reported in Refs.~\cite{Lees:2012kxa} and \cite{Aubert:2008bj}. Efficiencies range between $14\%$ and $28\%$ (see Table~\ref{tab:EffYieldBFSummary}).\\

In measurements of CP-violating asymmetries, the observed charge-specific raw event-yield asymmetries $\mathcal{A}$ are in general due to the combination of genuine CP-violating effects in the decay dynamics and instrumental asymmetries due to differences in interaction or reconstruction probabilities between particles and antiparticles. Such a combination is additive for small asymmetries,  $\mathcal{A}=\mathcal{A}_{\rm CP}+\mathcal{A}_{\rm det}$, with
\begin{equation*}
    \mathcal{A}_{\rm det}(X)=\frac{X-\overline{X}}{X+\overline{X}},
    \end{equation*}
where $X$ corresponds to a given final  state and $\overline{X}$ to its charge-conjugate.
Hence, observed raw charge-specific decay yields need be corrected for instrumental effects to determine the genuine CP-violating asymmetries.

\begin{figure}[htb]
 \centering
 \includegraphics[width=0.475\textwidth]{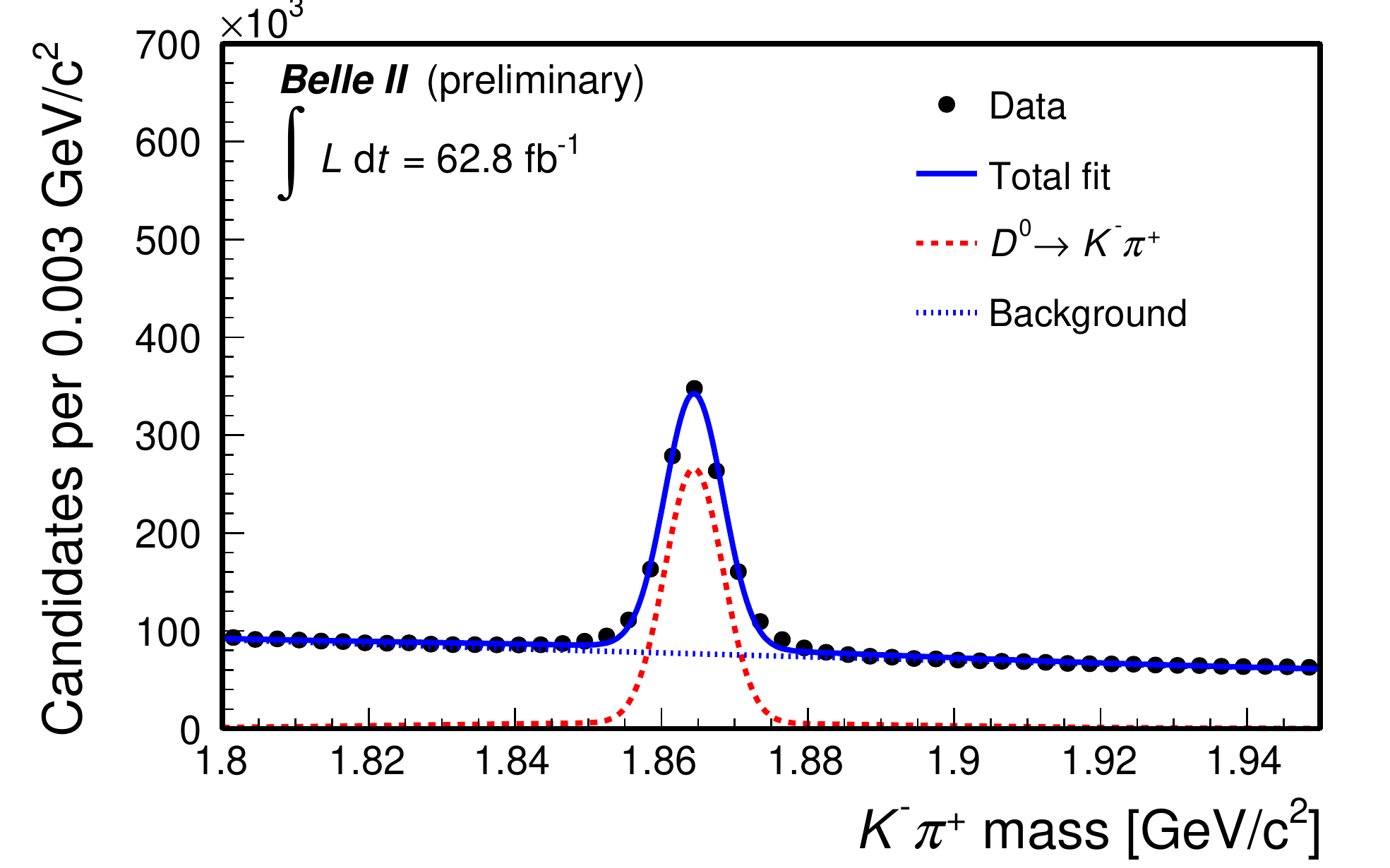}
 \includegraphics[width=0.475\textwidth]{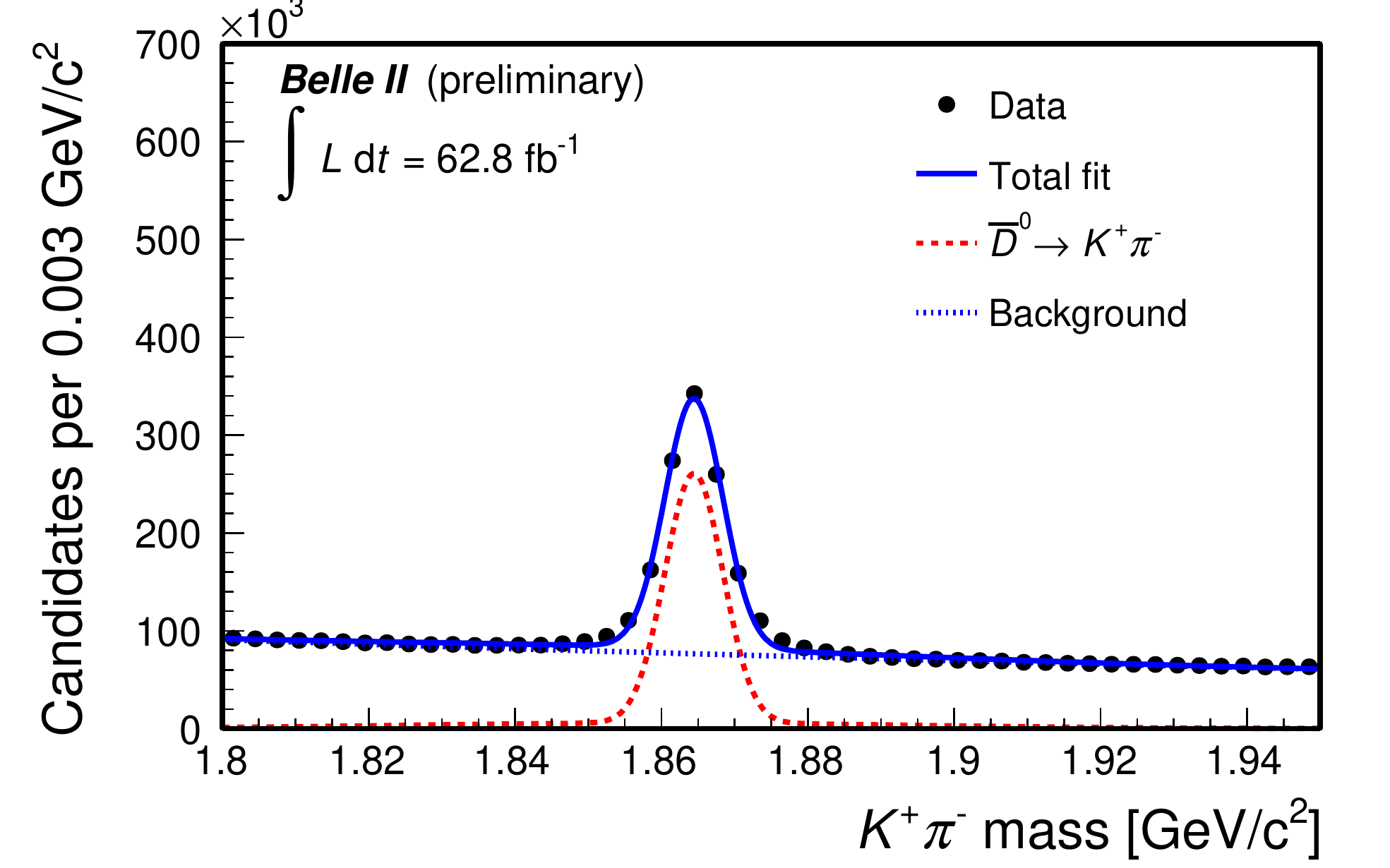}
 \caption{Distributions of $K^-\pi^+$ mass for (left) $D^0 \to K^-\pi^+$ and (right) $\overline{D}^0 \to  K^+\pi^-$ candidates reconstructed in 2019--2020 Belle~II data selected with an optimized continuum-suppression and kaon-enriching selection. The projection of an unbinned maximum likelihood fit is overlaid.}
 \label{fig:fit_invM_DtoKpi}
\end{figure}

We estimate the instrumental asymmetry $\mathcal{A}_{\rm det}(K\pi)$ associated with the reconstruction of $K^\pm\pi^\mp$ pairs by measuring the charge asymmetry in an abundant sample of $D^0 \to K^- \pi^+$ decays. For these decays, direct CP violation is expected to be smaller than 0.1\%~\cite{Zyla:2020zbs}. We therefore attribute any observed nonzero asymmetry to instrumental charge asymmetries. Figure~\ref{fig:fit_invM_DtoKpi} shows the $K^\pm\pi^\mp$-mass  distributions   for $D^0 \to K^-\pi^+$ and $\overline{D}^0 \to  K^+\pi^-$ candidates with fit projections overlaid. The resulting $K^\pm\pi^\mp$ asymmetry $\mathcal{A}_{\rm det}(K\pi)$ is directly subtracted from the raw measurements of charge-dependent yield asymmetry in $B^0 \to K^+\pi^-\pi^0$ decays to extract the corresponding CP-violating symmetry.\\

\begin{figure}[htb]
 \centering
 \includegraphics[width=0.475\textwidth]{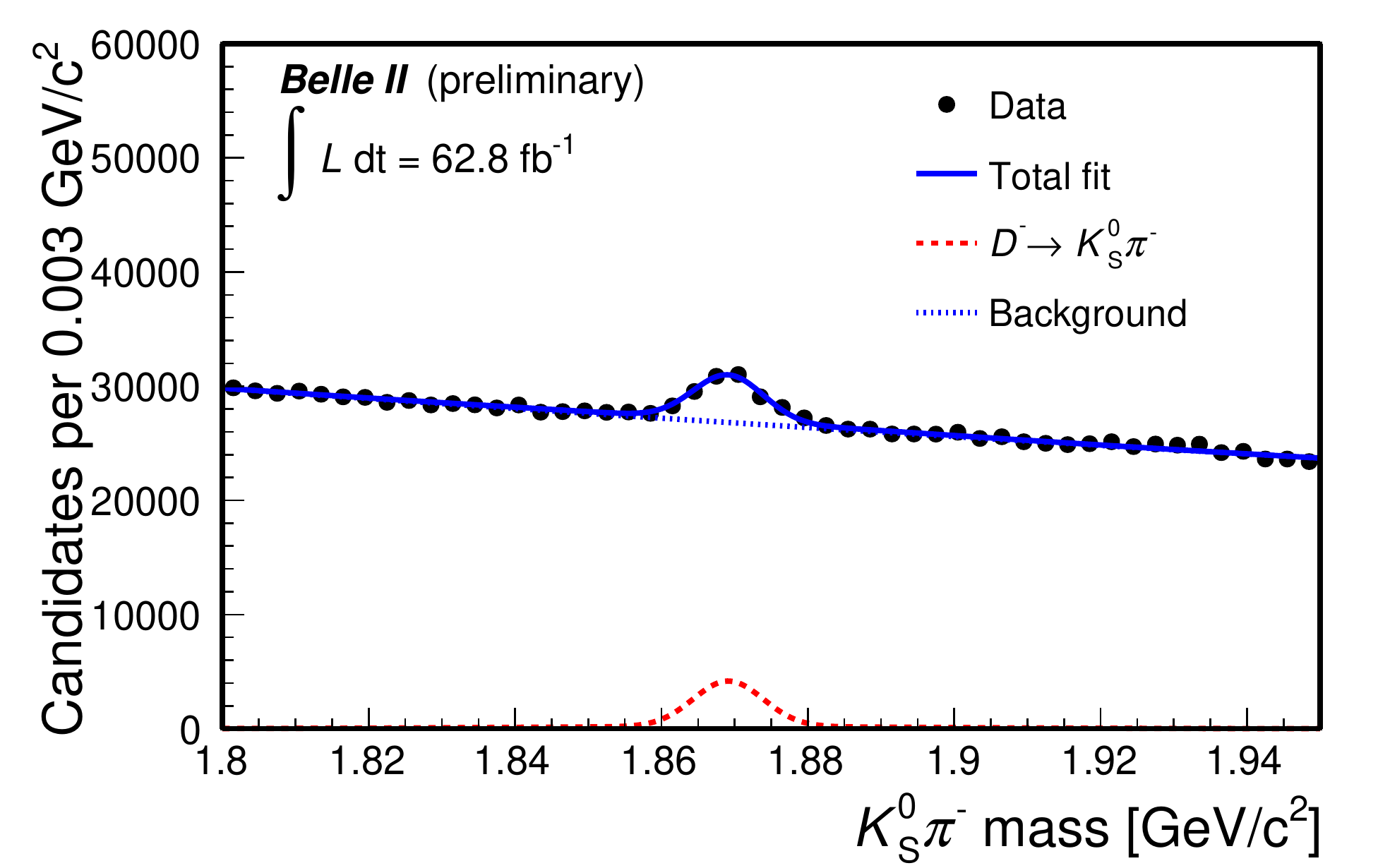}
 \includegraphics[width=0.475\textwidth]{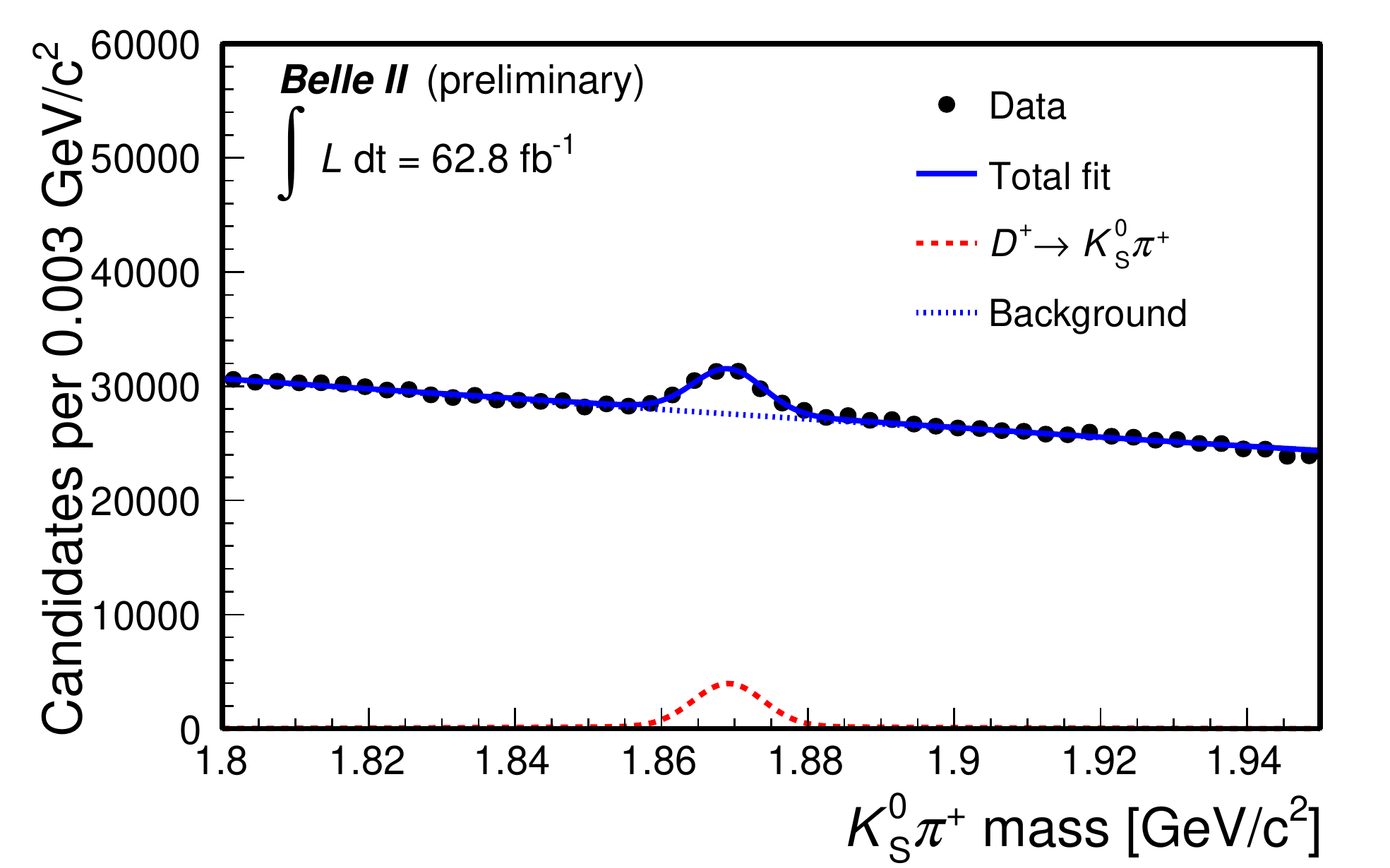}
 \caption{Distributions of $K^0_S\pi^+$ mass for (left) $D^- \to \PKzS\pi^-$ and (right) $D^+ \to  \PKzS\pi^+$ candidates reconstructed in 2019--2020 Belle~II data selected with an optimized continuum-suppression selection. Fit projections are overlaid.}
 \label{fig:fit_invM_DtoK0pi}
\end{figure}

For the $B^+ \to K^+\pi^-\pi^+$ and $B^+ \to K^+K^-K^+$ measurements, we correct for the instrumental asymmetry $\mathcal{A}_{\rm det}(K)$ related to charged kaon reconstruction. We determine this instrumental asymmetry by using the relationship \mbox{$\mathcal{A}_{\rm det}(K)=\mathcal{A}_{\rm det}(K\pi)-\mathcal{A}_{\rm det}(\PKzS\pi)+\mathcal{A}(\PKzS)$}. We obtain $\mathcal{A}_{\rm det}(\PKzS\pi)$ by measuring the yield asymmetry observed in an abundant sample of $D^+ \to \PKzS \pi^+$ decays~(Fig.~\ref{fig:fit_invM_DtoK0pi}), in which direct CP violation is expected to vanish. We estimate the component $\mathcal{A}(\PKzS)$~due to CP violation in neutral kaons by using the results obtained by the LHCb collaboration~\cite{LHCbInstr:2018}, which are consistent with previous assumptions at Belle~\cite{Ko:2010mk}. The resulting $K^\pm$ asymmetry is subtracted from the raw measurements of charge-dependent decay rates in $B^+ \to K^+K^-K^+$ and $B^+ \to K^+\pi^-\pi^+$ to extract the physics asymmetries. In each case, control channel selections are tuned to reproduce the kinematic conditions of the charmless final states that receive the corrections.  Table~\ref{tab:InstrChargeAsym} shows the resulting corrections.
\begin{table}[htb]
    \centering
\caption{Instrumental charge-asymmetries associated with $K^\pm\pi^\mp$, $\PKzS\pi^\pm$, and $K^\pm$  reconstruction, obtained using samples of $D^0 \to K^- \pi^+$ and $D^+ \to \PKzS \pi^+$ decays.}
\label{tab:InstrChargeAsym}
\begin{tabular}{l  c  c}
\hline\hline
Instrumental asymmetry & Value \\
\hline
$\mathcal{A}_{\rm det}(K^+\pi^-)$   & $-0.010 \pm 0.001$ \\
$\mathcal{A}_{\rm det}(\PKzS\pi^+)$   & $+0.026 \pm 0.019$ \\
$\mathcal{A}_{\rm det}(K^+)$   & $+0.017 \pm 0.019$ \\
\hline\hline
\end{tabular}
\end{table}
\clearpage
\section{Determination of branching fractions and CP-violating asymmetries}

We determine each branching fraction as 
\begin{equation*}
    \mathcal{B} = \frac{N}{\varepsilon\times 2\times N_{\PB\APB}},
\end{equation*}
where $N$ is the signal yield obtained from the fit, $\varepsilon$ is the reconstruction and selection efficiency, and $N_{\PB\APB}$ is the number of produced ${\PB\APB}$~pairs, corresponding to $35.8$~million for $\PBplus\PBminus$ and $33.9$~million for $\PBzero\APBzero$ pairs. We obtain $N_{\PB\APB}$ from the measured integrated luminosity, the \mbox{$\Pep\Pem\to\Upsilon(4{\rm S})$}~cross section~$(1.110 \pm 0.008)\,$nb~\cite{Bevan:2014iga}~(assuming that the $\Upsilon(4{\rm S})$ decays exclusively to ${\PB\APB}$~pairs), and the \mbox{$\Upsilon(4{\rm S})\to\PBzero\APBzero$}~branching fraction \mbox{$f^{00} = 0.487\pm 0.010\pm 0.008$}~\cite{Aubert:2005bq}.

The determination of CP-violating asymmetries is more straightforward because all factors that impact symmetrically bottom and antibottom rates cancel, and only flavor-specific yields and flavor-specific efficiency corrections are relevant.
\begin{table}[!ht]
    \centering
\caption{Summary of signal efficiencies $\varepsilon$, decay yields in 2019-2020 Belle~II data, and resulting branching fractions. Only statistical uncertainties are reported.}
    \label{tab:EffYieldBFSummary}
\begin{tabular}{l   r  r  r }
\hline\hline
Decay & \multicolumn{1}{c}{$\varepsilon\,[\%]$} & \multicolumn{1}{c}{Yield} & \multicolumn{1}{c}{$\mathcal{B}\,[10^{-6}]$} \\\hline
  $B^+ \to K^+ K^- K^+$   & $28.4\quad$ & $690\pm 30$        &	$35.8 \pm 1.6$ \\ 
  
  $B^+ \to K^+ \pi^- \pi^+$ & $18.6\quad$ & $843\pm 42$        &	$67.0 \pm 3.3$ \\
  
  $B^0 \to K^+ \pi^- \pi^0$ & $14.7\quad$ & $380\pm 35$        &	$38.1 \pm 3.5$ \\
   \hline\hline
\end{tabular}
\end{table}{}

\section{Systematic uncertainties}

We consider several sources of systematic uncertainties, assumed to be independent, and add in quadrature the corresponding uncertainties. An overview of the effects considered follows. A summary of the fractional size of systematic uncertainties is in  Tables~\ref{tab:SystematicsBF_overview} and \ref{tab:SystematicsAcp_overview}.

\subsection{Tracking efficiency} 
We assess a systematic uncertainty associated with possible data-simulation discrepancies in the reconstruction of charged particles~\cite{Bertacchi:2020eez}.
The tracking efficiency in data agrees with the value observed in simulation within a $0.91\%$ uncertainty, which we (linearly) add as a systematic uncertainty for each final-state charged particle.

\subsection{\Pgpz~reconstruction efficiency} 
We assess a systematic uncertainty associated with possible data-simulation discrepancies in the $\Pgpz$ reconstruction and selection using the decays \mbox{$B^0 \to D^{*-}(\to \overline{D}^0 (\to K^+ \pi^- \pi^0)\, \pi^-)\, \pi^+$} and \mbox{$B^0 \to D^{*-}(\to \overline{D}^0 (\to K^+ \pi^-)\, \pi^-)\, \pi^+$} where the selection of charged particle is identical and all distributions are weighted so that the $\pi^0$ momentum matches that in the $B^0 \to K^+ \pi^- \pi^0$ channel. We compare the yields obtained from fits to the $\Delta E$ distribution of reconstructed $\PB$~candidates and obtain a ratio between the $\Pgpz$ reconstruction efficiency in simulation and in data compatible with one. The $9.7\%$ uncertainty on this ratio is used as a systematic uncertainty.

\subsection{Particle-identification and continuum-suppression efficiencies} We evaluate possible data-simulation discrepancies in the particle identification and in the continuum-suppression distributions using the control channel \mbox{$\PBplus\to\APDzero(\to\PKp\Pgpm)\,\Pgpp$}. Selection efficiencies obtained in data and simulation agree within $0.7\%-1.3\%$ uncertainties (depending on the selection), which are taken as systematic uncertainties.

\subsection{Number of $\PB\APB$~pairs} We  assign  a  $1.4\%$  systematic  uncertainty on the number of $\PB\APB$~pairs,  which  includes  the uncertainty  on  cross-section,  integrated  luminosity~\cite{Abudinen:2019osb}, and  potential  shifts  from  the peak center-of-mass energy during the run periods.

\subsection{Signal modeling} 
Because we have empirical fit models for signal, we assess a systematic uncertainty associated with the model choice.
We use ensembles of simplified simulated experiments, in which the distribution for signal and background models are generated according to the default fitting model or to plausible alternative models.
We fit the composition of the simplified simulated samples using the same likelihood as for the data and use the difference between the means of the signal-yield distributions to determine a systematic uncertainty of $0.7\%-2.0\%$ for the branching-fraction measurement.

\subsection{Continuum background modeling} We apply the same procedure to assess the effect of possible continuum background mismodeling, obtaining uncertainties in the $0.1\%-1.3\%$ range for the branching-fraction measurement.

\subsection{Peaking and $\PB\APB$~background model} 
We apply the same procedure to assess the effect of $\PB\APB$ background mismodeling, obtaining uncertainties of typically $0.6\%-0.8\%$ for the branching-fraction measurement, and $0.005-0.009$ for the CP~asymmetry measurements.

\subsection{Instrumental asymmetries}
We consider the uncertainties on the values of  $\mathcal{A}_{\rm det}$~(Table~\ref{tab:InstrChargeAsym}) as systematic uncertainties due to instrumental asymmetry corrections in measurements of CP~asymmetries.

\begin{table}[h]
\caption{Summary of the (fractional) systematic uncertainties of the branching-fraction measurements.}
\label{tab:SystematicsBF_overview}
\centering
\footnotesize
\begin{tabular}{l c c c c c c c c}
\hline\hline
 Source &  $K^+ K^- K^+$ & $K^+ \pi^- \pi^+$ & $K^+ \pi^- \pi^0$\\
\hline
Tracking             &  2.73\% & 2.73\% & 1.82\% \\
 $\Pgpz$ efficiency      & - & - & 9.7\% \\
PID and continuum-supp. eff.    & 1.30\%  & 0.70\% & 0.65\% \\
 Signal efficiency & 0.2\% & 0.5\% & 1.6\% \\
 $N_{B\bar{B}}$      & 1.40\% & 1.40\%   & 1.40\% \\
  Signal model  & 1.9\%   & 0.72\% & 0.70\%\\
 Continuum bkg. model & 0.06\%   & 0.52\% & 1.30\%\\
 $\PB\APB$ bkg. model & 0.63\%   & 0.80\%  & 0.80\%\\
\hline
 Total  & 3.89\%   & 3.40\% & 10.25\%\\
\hline\hline
\end{tabular} 
\end{table}

\begin{table}[h]
\centering
\caption{Summary of (absolute) systematic uncertainties in the $\mathcal{A_{\rm CP}}$ measurements.}
\label{tab:SystematicsAcp_overview}
\footnotesize
\begin{tabular}{l c c c c c c}
\hline\hline
 Source & $K^+ K^- K^+$ & $K^+ \pi^- \pi^+$ & $K^+ \pi^- \pi^0$\\
\hline
 Signal model   & 0.002 & 0.005 & 0.006\\
 Pkg./$\PB\APB$/SCF background model & 0.005 & 0.006 & 0.009 \\
Instrumental asymmetry corrections  & 0.019 & 0.019 & 0.001 \\
\hline
 Total & 0.020 & 0.021 & 0.011 \\
\hline\hline
\end{tabular} 
\end{table}

\section{Results and summary}
\label{sec:summary}
We report on first measurements of  branching fractions and CP-violating charge asymmetries in charmless $B$ decays at Belle~II.  We use a sample of 2019 and 2020 data corresponding to $62.8\,\si{fb^{-1}}$ of integrated luminosity. We use simulation to devise optimized event selections. The $\Delta E$ and $M_{\rm bc}$ distributions of the resulting samples are fit to determine signal yields of approximately 690, 840, and 380 decays for the channels \mbox{$B^+ \to K^+K^-K^+$}, \mbox{$B^+ \to K^+\pi^-\pi^+$}, and \mbox{$B^0 \to K^+\pi^-\pi^0$}, respectively. Signal yields are corrected for efficiencies to obtain
\begin{center}
$\mathcal{B}(B^+ \to K^+K^-K^+) = [35.8 \pm 1.6(\rm stat) \pm 1.4 (\rm syst)]\times 10^{-6}$, 
\end{center}
\begin{center}
$\mathcal{B}(B^+ \to K^+\pi^-\pi^+) = [67.0 \pm 3.3 (\rm stat)\pm 2.3 (\rm syst)]\times 10^{-6}$,
\end{center}
\begin{center}
$\mathcal{B}(B^0 \to K^+\pi^-\pi^0) = [38.1 \pm 3.5 (\rm stat)\pm 3.9 (\rm syst)]\times 10^{-6}$, 
\end{center}
\begin{center}
$\mathcal{A}_{\rm CP}(B^+ \to K^+K^-K^+) = -0.103 \pm 0.042(\rm stat) \pm 0.020 (\rm syst)$,
\end{center}
\begin{center}
$\mathcal{A}_{\rm CP}(B^+ \to K^+\pi^-\pi^+) = -0.010 \pm 0.050 (\rm stat)\pm 0.021(\rm syst)$, and 
\end{center}
\begin{center}
$\mathcal{A}_{\rm CP}(B^0 \to K^+\pi^-\pi^0) = 0.207 \pm 0.088 (\rm stat)\pm 0.011(\rm syst)$.
\end{center}
These results are consistent with previous measurements and demonstrate detector performance comparable with the best Belle results, thus offering a reliable basis to assess projections for future reach.

\clearpage

\section*{Acknowledgments}

We thank the SuperKEKB group for the excellent operation of the
accelerator; the KEK cryogenics group for the efficient
operation of the solenoid; the KEK computer group for
on-site computing support; and the raw-data centers at
BNL, DESY, GridKa, IN2P3, and INFN for off-site computing support.
This work was supported by the following funding sources:
Science Committee of the Republic of Armenia Grant No. 20TTCG-1C010;
Australian Research Council and research grant Nos.
DP180102629, 
DP170102389, 
DP170102204, 
DP150103061, 
FT130100303, 
FT130100018,
and
FT120100745;
Austrian Federal Ministry of Education, Science and Research,
Austrian Science Fund No. P 31361-N36, and
Horizon 2020 ERC Starting Grant no. 947006 ``InterLeptons''; 
Natural Sciences and Engineering Research Council of Canada, Compute Canada and CANARIE;
Chinese Academy of Sciences and research grant No. QYZDJ-SSW-SLH011,
National Natural Science Foundation of China and research grant Nos.
11521505,
11575017,
11675166,
11761141009,
11705209,
and
11975076,
LiaoNing Revitalization Talents Program under contract No. XLYC1807135,
Shanghai Municipal Science and Technology Committee under contract No. 19ZR1403000,
Shanghai Pujiang Program under Grant No. 18PJ1401000,
and the CAS Center for Excellence in Particle Physics (CCEPP);
the Ministry of Education, Youth and Sports of the Czech Republic under Contract No.~LTT17020 and 
Charles University grants SVV 260448 and GAUK 404316;
European Research Council, 7th Framework PIEF-GA-2013-622527, 
Horizon 2020 ERC-Advanced Grants No. 267104 and 884719,
Horizon 2020 ERC-Consolidator Grant No. 819127,
Horizon 2020 Marie Sklodowska-Curie grant agreement No. 700525 `NIOBE,' 
and
Horizon 2020 Marie Sklodowska-Curie RISE project JENNIFER2 grant agreement No. 822070 (European grants);
L'Institut National de Physique Nucl\'{e}aire et de Physique des Particules (IN2P3) du CNRS (France);
BMBF, DFG, HGF, MPG, and AvH Foundation (Germany);
Department of Atomic Energy under Project Identification No. RTI 4002 and Department of Science and Technology (India);
Israel Science Foundation grant No. 2476/17,
United States-Israel Binational Science Foundation grant No. 2016113, and
Israel Ministry of Science grant No. 3-16543;
Istituto Nazionale di Fisica Nucleare and the research grants BELLE2;
Japan Society for the Promotion of Science,  Grant-in-Aid for Scientific Research grant Nos.
16H03968, 
16H03993, 
16H06492,
16K05323, 
17H01133, 
17H05405, 
18K03621, 
18H03710, 
18H05226,
19H00682, 
26220706,
and
26400255,
the National Institute of Informatics, and Science Information NETwork 5 (SINET5), 
and
the Ministry of Education, Culture, Sports, Science, and Technology (MEXT) of Japan;  
National Research Foundation (NRF) of Korea Grant Nos.
2016R1\-D1A1B\-01010135,
2016R1\-D1A1B\-02012900,
2018R1\-A2B\-3003643,
2018R1\-A6A1A\-06024970,
2018R1\-D1A1B\-07047294,
2019K1\-A3A7A\-09033840,
and
2019R1\-I1A3A\-01058933,
Radiation Science Research Institute,
Foreign Large-size Research Facility Application Supporting project,
the Global Science Experimental Data Hub Center of the Korea Institute of Science and Technology Information
and
KREONET/GLORIAD;
Universiti Malaya RU grant, Akademi Sains Malaysia and Ministry of Education Malaysia;
Frontiers of Science Program contracts
FOINS-296,
CB-221329,
CB-236394,
CB-254409,
and
CB-180023, and SEP-CINVESTAV research grant 237 (Mexico);
the Polish Ministry of Science and Higher Education and the National Science Center;
the Ministry of Science and Higher Education of the Russian Federation,
Agreement 14.W03.31.0026;
University of Tabuk research grants
S-0256-1438 and S-0280-1439 (Saudi Arabia);
Slovenian Research Agency and research grant Nos.
J1-9124
and
P1-0135; 
Agencia Estatal de Investigacion, Spain grant Nos.
FPA2014-55613-P
and
FPA2017-84445-P,
and
CIDEGENT/2018/020 of Generalitat Valenciana;
Ministry of Science and Technology and research grant Nos.
MOST106-2112-M-002-005-MY3
and
MOST107-2119-M-002-035-MY3, 
and the Ministry of Education (Taiwan);
Thailand Center of Excellence in Physics;
TUBITAK ULAKBIM (Turkey);
Ministry of Education and Science of Ukraine;
the US National Science Foundation and research grant Nos.
PHY-1807007 
and
PHY-1913789, 
and the US Department of Energy and research grant Nos.
DE-AC06-76RLO1830, 
DE-SC0007983, 
DE-SC0009824, 
DE-SC0009973, 
DE-SC0010073, 
DE-SC0010118, 
DE-SC0010504, 
DE-SC0011784, 
DE-SC0012704, 
DE-SC0021274; 
and
the Vietnam Academy of Science and Technology (VAST) under grant DL0000.05/21-23.

\bibliography{belle2}

\providecommand{\href}[2]{#2}\begingroup\raggedright\begin{thebibliography}{10}

\bibitem{Lees:2012kxa}
J.~P. Lees {\it et al.} , {(BaBar Collaboration)}, { {Study of CP violation in
  Dalitz-plot analyses of $B^0 \to K^+K^-K^0_S$, $B^+ \to K^+K^-K^+$, and $B^+
  \to K^0_S K^0_S K^+$}\/},
  \href{http://dx.doi.org/10.1103/PhysRevD.85.112010}{\color{blue}Phys. Rev. D
  {\bf 85} (2012)  112010},
  \href{http://arxiv.org/abs/1201.5897}{{\color{blue} \tt arXiv:1201.5897}}.

\bibitem{Aubert:2008bj}
B.~Aubert {\it et al.} , {(BaBar Collaboration)}, { {Evidence for Direct CP
  Violation from Dalitz-plot analysis of $B^\pm \to K^\pm \pi^\mp \pi^\pm$}\/},
   \href{http://dx.doi.org/10.1103/PhysRevD.78.012004}{\color{blue}Phys. Rev. D
  {\bf 78} (2008)  012004},
  \href{http://arxiv.org/abs/0803.4451}{{\color{blue} \tt arXiv:0803.4451
  [hep-ex]}}.

\bibitem{PhysRevLett.112.011801}
R.~Aaij {\it et al.} , {LHCb Collaboration}, { Measurement of $CP$ Violation in
  the Phase Space of
  ${B}^{\ifmmode\pm\else\textpm\fi{}}\ensuremath{\rightarrow}{K}^{+}{K}^{\ensuremath{-}}{\ensuremath{\pi}}^{\ifmmode\pm\else\textpm\fi{}}$
  and
  ${B}^{\ifmmode\pm\else\textpm\fi{}}\ensuremath{\rightarrow}{\ensuremath{\pi}}^{+}{\ensuremath{\pi}}^{\ensuremath{-}}{\ensuremath{\pi}}^{\ifmmode\pm\else\textpm\fi{}}$
  Decays\/},
  \href{http://dx.doi.org/10.1103/PhysRevLett.112.011801}{\color{blue}Phys.
  Rev. Lett. {\bf 112} (2014)  011801}.

\bibitem{JHEP10(2017)117}
R.~Klein {\it et al.} , { CP violation in multibody $B$ decays from QCD
  factorization.\/},
  \href{http://dx.doi.org/10.1007/JHEP10(2017)117}{\color{blue}J. High Energ.
  Phys. {\bf 117} (2017)  }.

\bibitem{PhysRevD.94.094015}
H.~Cheng, C.-K. Chua, and Z.-Q. Zhang, { Direct $CP$ violation in charmless
  three-body decays of $B$ mesons\/},
  \href{http://dx.doi.org/10.1103/PhysRevD.94.094015}{\color{blue}Phys. Rev. D
  {\bf 94} (2016)  094015}.

\bibitem{Abudinen:2019osb}
F.~Abudin\'en {\it et al.} {(Belle II Collaboration)}, { {Measurement of the
  integrated luminosity of the Phase 2 data of the Belle II experiment}\/},
  \href{http://dx.doi.org/10.1088/1674-1137/44/2/021001}{\color{blue}Chin.
  Phys. C {\bf 44} (2020) no.~2, 021001}.

\bibitem{Benedikt:2019}
B.~Wach{\;(Belle~II Collaboration)}, { {First charmless $B$ signal
  reconstruction in Belle II}\/},
  \href{https://docs.belle2.org/record/1665?ln=en,
  }{\color{blue}BELLE2-NOTE-PL-2019-25}.

\bibitem{CharmlessMoriond:2020}
F.~Abudin{\'e}n {\it et al.} {(Belle II Collaboration)}, { {Charmless $B$-decay
  reconstruction in 2019 Belle~II data}\/},
  \href{http://arxiv.org/abs/2005.13559}{{\color{blue} \tt arXiv:2005.13559}}.

\bibitem{Kou:2018nap}
W.~Altmannshofer {\it et al.} {(Belle~II Collaboration)}, { {The Belle II
  Physics Book}\/},  \href{http://dx.doi.org/10.1093/ptep/ptz106,
  10.1093/ptep/ptaa008}{\color{blue}PTEP {\bf 2019} (2019) no.~12, 123C01}.

\bibitem{Abe:2010sj}
T.~Abe {\it et al.} {(Belle II Collaboration)}, { {Belle II Technical Design
  Report}\/},   \href{http://arxiv.org/abs/1011.0352}{{\color{blue} \tt
  arXiv:1011.0352}}.

\bibitem{Akai:2018mbz}
K.~Akai {\it et al.} {(SuperKEKB)}, { {SuperKEKB Collider}\/},
  \href{http://dx.doi.org/10.1016/j.nima.2018.08.017}{\color{blue}Nucl.\
  Instrum.\ Meth.\ A {\bf 907} (2018)  188--199}.

\bibitem{Ryd:2005zz}
A.~T. Ryd {\it et al.} { {EvtGen: A Monte Carlo Generator for B-Physics}\/},
  EVTGEN-V00-11-07 (2005)  .

\bibitem{Kuhr:2018lps}
T.~Kuhr {\it et al.} { {The Belle II Core Software}\/},
  \href{http://dx.doi.org/10.1007/s41781-018-0017-9}{\color{blue}Comput. Softw.
  Big Sci. {\bf 3} (2019) no.~1, 1}.

\bibitem{Zyla:2020zbs}
P.~Zyla {\it et al.} , {Particle Data Group}, { {Review of Particle
  Physics}\/},  \href{http://dx.doi.org/10.1093/ptep/ptaa104}{\color{blue}PTEP
  {\bf 2020} (2020) no.~8, 083C01}.

\bibitem{Abudinen:2018}
F.~Abudin{\'e}n{, Ph.D. Thesis}, { {Development of a $\PBzero$~flavor tagger
  and performance study of a novel time-dependent $\CP$ analysis of the decay
  $\PBzero\to\Pgpz\Pgpz$ at Belle~II, Ludwig Maximilian University of Munich
  (2018)}\/},  \href{https://docs.belle2.org/record/1215?ln=en,
  }{\color{blue}BELLE2-PTHESIS-2018-003}.

\bibitem{Skwarnicki:1986xj}
T.~Skwarnicki{, Ph.D. Thesis}, { {A study of the radiative CASCADE transitions
  between the Upsilon-Prime and Upsilon resonances, Institute of Nuclear
  Physics, Krakow (1986)}\/},
  \href{https://inspirehep.net/files/e31108fa63ba2754e30038cccdf9e3a7,
  }{\color{blue}DESY-F31-86-02}.

\bibitem{ALBRECHT1990278}
H.~Albrecht {\it et al.} , { Search for hadronic b→u decays\/},
  \href{http://dx.doi.org/https://doi.org/10.1016/0370-2693(90)91293-K}{\color{blue}Physics
  Letters B {\bf 241} (1990) no.~2, 278--282}.

\bibitem{LHCbInstr:2018}
A.~Davis {\it et al.} {(LHCb Collaboration)}, { {Measurement of the
  instrumental asymmetry for \mbox{$K^-\pi^+$}-pairs at LHCb in Run~2}\/},
  \href{https://cds.cern.ch/record/2310213/files/LHCb-PUB-2018-004.pdf,
  }{\color{blue}LHCb-PUB-2018-004}.

\bibitem{Ko:2010mk}
B.~R. Ko, E.~Won, B.~Golob, and P.~Pakhlov, { {Effect of nuclear interactions
  of neutral kaons on CP asymmetry measurements}\/},
  \href{http://dx.doi.org/10.1103/PhysRevD.84.111501}{\color{blue}Phys. Rev. D
  {\bf 84} (2011)  111501},
  \href{http://arxiv.org/abs/1006.1938}{{\color{blue} \tt arXiv:1006.1938
  [hep-ex]}}.

\bibitem{Bevan:2014iga}
A.~J. Bevan {\it et al.} {(Belle and BaBar Collaborations)}, { {The Physics of
  the $B$ Factories}\/},
  \href{http://dx.doi.org/10.1140/epjc/s10052-014-3026-9}{\color{blue}Eur.
  Phys. J. {\bf C74} (2014)  3026}.

\bibitem{Aubert:2005bq}
B.~Aubert {\it et al.} {(BaBar Collaboration)}, { {Measurement of the branching
  fraction of \mbox{$\Upsilon(4{\rm S}) \to B^0 \overline{B}^0$}}\/},
  \href{http://dx.doi.org/10.1103/PhysRevLett.95.042001}{\color{blue}Phys. Rev.
  Lett. {\bf 95} (2005)  042001}.

\bibitem{Bertacchi:2020eez}
V.~Bertacchi {\it et al.} {(Belle II tracking)}, { {Track Finding at Belle
  II}\/},   \href{http://arxiv.org/abs/2003.12466}{{\color{blue} \tt
  arXiv:2003.12466}}.

\end{thebibliography}\endgroup
\bibliographystyle{belle2-note}

\end{document}